\documentclass[
superscriptaddress,
aip,apl,
reprint,
twocolumn,
amsmath,amssymb,showpacs
]{revtex4-2}
\usepackage{titlesec}
\usepackage{newtxtext,newtxmath}
\usepackage{physics}
\usepackage{graphicx}
\usepackage{dsfont}
\usepackage{xcolor}
\definecolor{urlc}{RGB}{58,105,157}
\usepackage[
colorlinks=true,
urlcolor=urlc,
linkcolor=urlc,
citecolor=urlc
]{hyperref}
\titleformat{\section}{\large\sffamily\bfseries}{\thesection}{}{}
\titleformat{\subsection}[runin]{\sffamily\bfseries}{\thesubsection}{}{}
\titlespacing*{\section}{0pt}{3ex}{0ex}
\titlespacing*{\subsection}{0pt}{2ex}{1ex}

\newcommand{\ii}{\mathrm{i}}
\newcommand{\ee}{\mathrm{e}}
\begin{document}

\title{Detection of arbitrary quantum correlations via synthesized quantum channels}

\author{Ze Wu}%
\thanks{These authors contribute equally}
\affiliation{
CAS Key Laboratory of Microscale Magnetic Resonance and School of Physical Sciences, University of Science and Technology of China, Hefei 230026, China}
\affiliation{
CAS Center for Excellence in Quantum Information and Quantum Physics, University of Science and Technology of China, Hefei 230026, China}

\author{Ping Wang}%
\thanks{These authors contribute equally}
\affiliation{
College of Education for the future, Beijing Normal University at Zhuhai (BNU Zhuhai), Zhuhai, China}
\affiliation{
Department of Physics, The Chinese University of Hong Kong, Shatin, New Territories, Hong Kong, China}

\author{Tianyun Wang}
\affiliation{
CAS Key Laboratory of Microscale Magnetic Resonance and School of Physical Sciences, University of Science and Technology of China, Hefei 230026, China}
\affiliation{
CAS Center for Excellence in Quantum Information and Quantum Physics, University of Science and Technology of China, Hefei 230026, China}

\author{Yuchen Li}
\affiliation{
CAS Key Laboratory of Microscale Magnetic Resonance and School of Physical Sciences, University of Science and Technology of China, Hefei 230026, China}
\affiliation{
CAS Center for Excellence in Quantum Information and Quantum Physics, University of Science and Technology of China, Hefei 230026, China}

\author{Ran Liu}
\affiliation{
CAS Key Laboratory of Microscale Magnetic Resonance and School of Physical Sciences, University of Science and Technology of China, Hefei 230026, China}
\affiliation{
CAS Center for Excellence in Quantum Information and Quantum Physics, University of Science and Technology of China, Hefei 230026, China}

\author{Yuquan Chen}
\affiliation{
CAS Key Laboratory of Microscale Magnetic Resonance and School of Physical Sciences, University of Science and Technology of China, Hefei 230026, China}
\affiliation{
CAS Center for Excellence in Quantum Information and Quantum Physics, University of Science and Technology of China, Hefei 230026, China}

\author{Xinhua Peng}%
\email{xhpeng@ustc.edu.cn}
\affiliation{
CAS Key Laboratory of Microscale Magnetic Resonance and School of Physical Sciences, University of Science and Technology of China, Hefei 230026, China}
\affiliation{
CAS Center for Excellence in Quantum Information and Quantum Physics, University of Science and Technology of China, Hefei 230026, China}

\author{Ren-Bao Liu}%
\email{rbliu@cuhk.edu.hk}
\affiliation{
Department of Physics, The Chinese University of Hong Kong, Shatin, New Territories, Hong Kong, China}
\affiliation{
Centre for Quantum Coherence, The Chinese University of Hong Kong, Shatin, New Territories, Hong Kong, China}
\affiliation{
The Hong Kong Institute of Quantum Information Science and Technology, The Chinese University of Hong Kong, Shatin, New Territories, Hong Kong, China}

\author{Jiangfeng Du}%
\affiliation{
CAS Key Laboratory of Microscale Magnetic Resonance and School of Physical Sciences, University of Science and Technology of China, Hefei 230026, China}
\affiliation{
CAS Center for Excellence in Quantum Information and Quantum Physics, University of Science and Technology of China, Hefei 230026, China}
 
\date{\today}

\begin{abstract}
    Quantum correlations are key information about the structures and dynamics of quantum many-body systems. There are many types of high-order quantum correlations with different time orderings, but only a few of them are accessible to the existing detection methods. Recently, a quantum-sensing approach based on sequential weak measurement was proposed to selectively extract arbitrary types of correlations. However, its experimental implementation is still elusive. Here we demonstrate the extraction of arbitrary types of quantum correlations. We generalized the original weak measurement scheme to a protocol using synthesized quantum channels, which can be applied to more universal scenarios including both single and ensemble quantum systems. In this quantum channel method, various controls on the sensors are superimposed to select the sensor-target evolution along a specific path for measuring a desired quantum correlation. Using the versatility of nuclear magnetic resonance techniques, we successfully extract the second- and fourth-order correlations of a nuclear-spin target by another nuclear-spin sensor. The full characterization of quantum correlations provides a new tool for understanding quantum many-body systems, exploring fundamental quantum physics, and developing quantum technologies.
\end{abstract}

\maketitle

\begin{figure*}
    \includegraphics[scale=1]{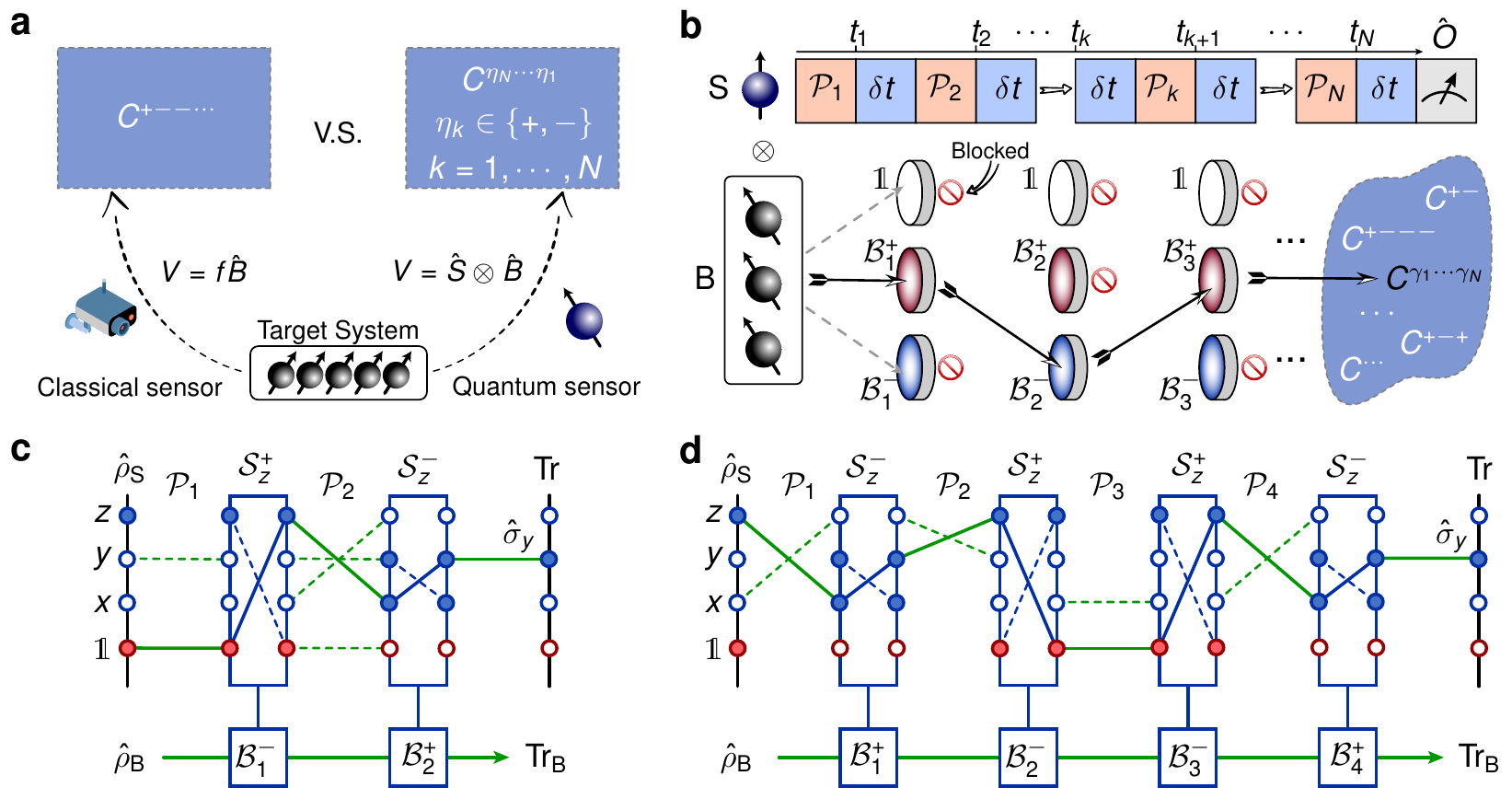}
    \caption{\textbf{Schematics of the measurement protocol for arbitrary quantum correlations.} \textbf{a}, Quantum sensors vs. classical sensors for correlation measurements. Only one type of correlations is accessible to a classical probe while arbitrary types of correlations can be extracted by a quantum sensor. \textbf{b}, General protocol to extract arbitrary types of quantum correlations with synthesized quantum channels on a quantum sensor. \textbf{c}, Diagram representation of a designed quantum channel for measuring the second-order correlation $ C^{+-} $. \textbf{d}, Diagram representation of a designed quantum channel for measuring the four-order correlation $ C^{+--+} $. The initial state $ \hat{\rho}_{\mathrm{S}} $ and observable $ \hat{\sigma}_{y} $ are represented by the four-vectors. The blue solid lines inside the rectangles denote the transfer paths of $ \mathcal{S}_{z}^{\pm} $ (e.g. the two lines in the rectangle of $ \mathcal{S}_{z}^{-} $ represent the transfer paths like $ \sigma_x\leftrightarrow \sigma_y $), and the green lines denote the transfer paths of the quantum channels $ \mathcal{P}_{k} $. The connected paths denoted by solid lines represent the selected ones to measure the desired correlations. Those denoted by the dashed lines are disconnected to block the unwanted correlations.}\label{Fig1}
\end{figure*}

\section*{Introduction }
Correlations of physical quantities are key to understanding quantum many-body physics\cite{Douglas2015,SchweiglerNature2017,Rubio2019}, nonlinear optics\cite{DorfmanRMP2016}, solid-state nuclear magnetic resonance (NMR)\cite{Alvarez2010,Alvarez2015} and open quantum systems\cite{Prokof2000,Vega2017,YangRPP2017,Gasbarri2018}, and are relevant to some quantum-enhanced technologies\cite{DegenRMP2017,Slichterbook2013,Mukamelbook1999}. Second-order correlations\cite{Clerk2010,SinitsynRPP2016} are the underlying physical quantities measured in a broad range of fields such as linear optics\cite{Carolan2015}, transport\cite{Mahan1987}, thermodynamics\cite{Vinjanampathy2016}, neutron scattering\cite{Schofield1960}, and are directly extracted in various quantum systems such as single solid impurities\cite{LaraouiNC2013,SimonScience2017,BossScience2017,PfenderNC2019,DegenNature2019}, quantum dots\cite{CrookerPRL2010,LiPRL2012} and superconductor qubits\cite{KuhlmannNP2013}. Recently, it is indicated that high-order correlations are relevant to mesoscopic quantum many-body systems\cite{Norris2016,SchweiglerNature2017,SchweiglerNP2021}. How to systematically characterize these correlations then plays a central role in investigation of various quantum systems\cite{DorfmanRMP2016,WangPRL2019}.

In general, the dynamics of a quantum system is determined by the correlations $ C^{\eta_{N}\cdots\eta_{1}}=\Tr_{\mathrm{B}}\left(\mathcal{B}_{N}^{\eta_{N}}\cdots\mathcal{B}_{2}^{\eta_{2}}\mathcal{B}_{1}^{\eta_{1}}\hat{\rho}\right) $, where $ \hat{\rho} $ is the initial state of the system and $ \eta_{k}=\pm 1 $. The super-operators of the commutator and anti-commutator at time $t_k$ (with $t_1\le t_2\le \cdots t_N$) are defined as $ \mathcal{B}_{k}^{-}\hat{\rho}\equiv-\ii[\hat{B}\left(t_{k}\right)\hat{\rho}-\hat{\rho}\hat{B}\left(t_{k}\right)]/2 $ and $ \mathcal{B}_{k}^{+}\hat{\rho}\equiv[\hat{B}\left(t_{k}\right)\hat{\rho}+\hat{\rho}\hat{B}\left(t_{k}\right)]/2 $, respectively. The quantum quantities $ \hat{B}\left(t_{k}\right) $ at different time $ t_{k} $ in general do not commute, i.e., $ \hat{B}\left(t_{i}\right)\hat{B}\left(t_{j}\right)\neq\hat{B}\left(t_{j}\right)\hat{B}\left(t_{i}\right) $ when $ i\neq j $. Therefore, the high-order quantum correlations have a rich structure resulting from different orderings\cite{Keldysh1965,Chou1985,Sieberer2016}. There are $ 2^{N-1} $ inequivalent correlations corresponding to different nesting of commutators and anti-commutators in time order\cite{Gasbarri2018,WangPRL2019,WangCPL2021}. Among these numerous correlations, several special forms have been widely investigated and play significant roles in many subjects. For example, the nonlinear spectroscopy\cite{Mukamelbook1999}, in which the system is detected by a classical sensor, measures only one type of time-ordered correlations, namely, $ C^{+--\cdots -}=\Tr_{\mathrm{B}}\left(\mathcal{B}_{N}^{+}\cdots\mathcal{B}_{2}^{-}\mathcal{B}_{1}^{-}\hat{\rho}_{\mathrm{B}}\right) $. Another widely used tool is the noise spectroscopy\cite{ZapasskiiPRL2013,YangRPP2017}, in which the target system is approximated as a classical stochastic noise field, and it usually extracts the symmetric correlations like $ C^{++\cdots +}=\Tr_{\mathrm{B}}\left(\mathcal{B}_{N}^{+}\cdots\mathcal{B}_{2}^{+}\mathcal{B}_{1}^{+}\hat{\rho}_{\mathrm{B}}\right) $. 
As shown in Fig.~\ref{Fig1}a, using the quantum sensors to detect the target systems is necessary for the full extraction of the quantum correlations, but few studies on this subject are carried out. Recently, it is proposed theoretically that the sequential weak measurements via a quantum sensor can extract arbitrary-order correlations of a quantum bath\cite{WangPRL2019}. By preparing the initial state and choosing a certain measurement basis of the sensor, one can pre- and post-select the coupling between the sensor and the target system to access arbitrary types and orders of correlations of the target system. A very recent experimental work also used the sequential weak measurement to obtain the mixed signals of the fourth-order correlations of single nuclear spin\cite{JonasarXiv2021}. Until now, selective detection of arbitrary types of correlations has not yet been realized in experiments.


Here we demonstrate the extraction of arbitrary types of quantum correlations. In stead of the original weak measurement scheme\cite{WangPRL2019}, we propose a more general protocol using synthesized quantum channels to select the coupled evolution of the quantum sensor and target system along a specific path, leading to a desired correlation detected by the final measurement. This quantum channel scheme is universal, applicable to both single and ensemble quantum systems. Using the versatility of NMR techniques, which are a powerful tool for studying quantum many-body physics\cite{Alvarez2015,Abragam1982} and correlation measurements\cite{Li2017,Niknam2020,Niknam2021}, we demonstrate the scheme using nuclear-spin targets and sensors. We extract the second-order correlation $ C^{+-} $ and fourth-order correlation $ C^{+--+} $. It is expected that the measurement protocol for arbitrary quantum correlations will provide an essential tool for studying quantum many-body physics and finding the applications in quantum technologies.

\begin{figure}
    \centering
    \includegraphics[scale=0.95]{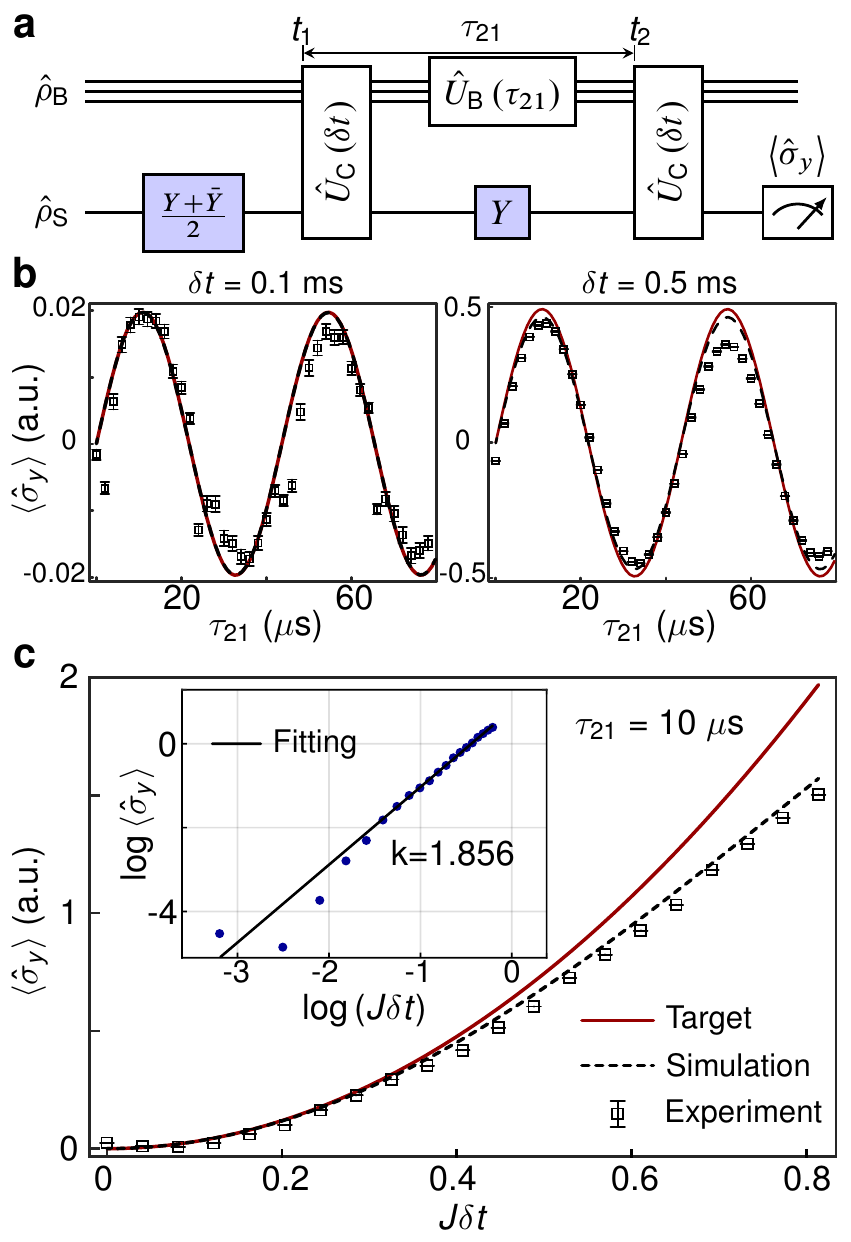}
    \caption{\textbf{Measurement of the second-order correlations $ C^{+-} $.}
    \textbf{a}, Experimental diagram for measuring second-order quantum correlation $ C^{+-} $. $ \hat{U}_{\rm C}(\delta t) = \ee^{-\ii \hat{V}\delta t} $ is the coupled evolution for the case of pure dephasing. $ \hat{U}_{\mathrm{B}}(\tau_{21}) $ is the free evolution of the quantum target under its Hamiltonian $\hat{H}_B$ for time $\tau_{21} = t_2- t_1$. The two synthesized quantum channels are $ \left(Y+\bar{Y}\right)/2 = [\mathcal{R}_{y}\left(\pi/2\right) + \mathcal{R}_{-y}\left(\pi/2\right)]/2 $ and $Y = \mathcal{R}_{y}\left(\pi/2\right)$. \textbf{b}, Measured values of $ \left<\hat{\sigma}_{y}^{\rm C}\right> $ versus the time interval $ \tau_{21} $ for $ \delta t=0.1 $ ms (left graph) and $ \delta t=0.5 $ (right graph). \textbf{c}, Measured values of $ \left<\hat{\sigma}_{y}^{\rm C}\right> $ versus $ J\delta t $ with a fixed interval $ \tau_{21}=10 $ $ \mu $s. The inset shows the linear fitting of the logarithmic data. In \textbf{b} and \textbf{c}, the red solid lines denote the theoretical curves of the target signals ($ \delta t^2 C^{+-} $), the black dashed lines denote the numerical simulations of $ S_{2} $ in equation~\eqref{methods2th} and the error bars are given based on the fitting errors of experimental spectra (see Supplementary Note 3).}\label{Fig2}
\end{figure}
\section*{Scheme}
\begin{figure*}
  \centering
  \includegraphics[scale=1]{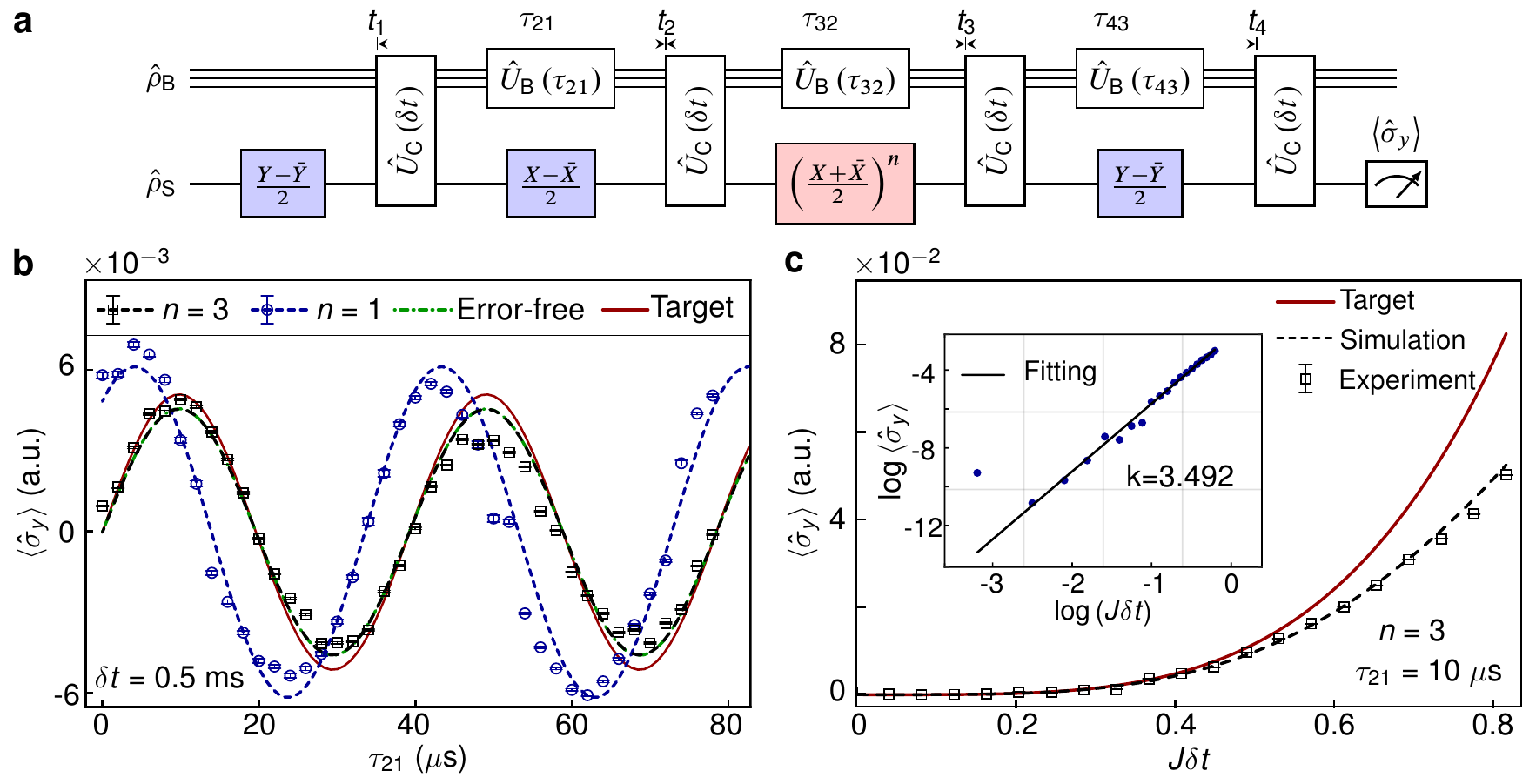}
  \caption{\textbf{Measurement of the fourth-order correlations $ C^{+--+} $.} \textbf{a}, Experimental diagram for measuring the fourth-order correlation $ C^{+--+} $. The symbols are the same as Figure 2a. $ X/\bar{X}=\mathcal{R}_{\pm x}(\pi/2) $ and $ Y/\bar{Y}=\mathcal{R}_{\pm y}(\pi/2) $. The quantum channel $ [(X +\bar{X})/2]^n$ is designed to be robust to the pulse errors by repeating the quantum channel $(X +\bar{X})/2$ for $ n $ times. \textbf{b}, Measured values of $\left<\hat{\sigma}_{y}^{\rm C}\right> $ versus $ \tau_{21} $ for $ n=1 $ (blue scatters) and $ n=3 $ (black scatters) when $ \delta t=0.5 $ ms and $ \tau_{32}=\tau_{43}=10 $ $\mu$s. The green dash-dotted line denotes the ideal simulated signal. \textbf{c}, Measured values of $ \left<\hat{\sigma}_{y}^{\rm C}\right> $ versus $ J\delta t $ for the fixed $ \tau_{21}=10 $ $ \mu $s. The inset is the linear fitting of the logarithmic data. In \textbf{b} and \textbf{c}, the red solid lines denote the theoretical curves of the target signals ($ \delta t^4 p_{\rm C}C^{+--+} $). The black and blue dashed lines denote the corresponding numerical simulations of $ S_{4} $ for $ n=1 $ and $ n=3 $ with $\pi/2$ pulse error. The error bars are given based on the fitting errors of experimental spectra.}\label{Fig3}
\end{figure*}
Consider a quantum sensor S transiently coupled to a quantum many-body target B at time $ t_{k} $ for a short time $ \delta t $, the state evolution is governed by the interaction Hamiltonian $ \hat{V}=\hat{S}\otimes\hat{B}(t_{k}) $. Then the corresponding Liouville equation $ \partial_{t}\hat{\rho}\left(t\right)=-\ii\left[\hat{V},\hat{\rho}\left(t\right)\right] $ has the first-order approximation\cite{Breuer2002}
\begin{equation}
    \hat{\rho}\left(t_{k}+\delta t\right)\approx\left[\mathds{1}+2\left(\mathcal{S}^{-}\otimes\mathcal{B}_{k}^{+}+\mathcal{S}^{+}\otimes\mathcal{B}_{k}^{-}\right)\delta t\right]\hat{\rho}\left(t_{k}\right),
\end{equation}
where $ \mathcal{S}^{\pm}/\mathcal{B}_{k}^{\pm} $ are the super-operators as defined before and $ \hat{B}_{k}=\hat{B}\left(t_{k}\right) $. Suppose the sensor and target are initially separable, i.e., $ \hat{\rho}\left(0\right)=\hat{\rho}_{\mathrm{S}}\otimes\hat{\rho}_{\mathrm{B}} $. After passing through the $ N $ transient interactions successively at $ t_{1}\leq t_{2}\leq\cdots\leq t_{N} $, the final state after $ t_{N} $ becomes
\begin{equation}
    \hat{\rho}_{\rm f}\approx\sum_{\eta_{k}\in\left\{\pm,0\right\}}\mathcal{T}\left[\left(2\delta t\right)^{\Theta}\prod_{k=1}^{N}\left(\mathcal{S}^{\overline{\eta}_{k}}\otimes\mathcal{B}_{k}^{\eta_{k}}\right)\right]\hat{\rho}_{\mathrm{S}}\otimes\hat{\rho}_{\mathrm{B}},\label{state}
\end{equation}
where $ \mathcal{T} $ is the time-ordering operator and $ \Theta=\sum_{k=1}^{N}\left|\eta_{k}\right| $. Here $ \eta_{k}\in\left\{\pm,0\right\} $ and we define $ \mathcal{S}^{0}=\mathcal{B}_{k}^{0}\equiv\mathds{1}, k=1,\cdots,N $. $ \overline{\eta}_{k}=-\eta_{k} $, which means $ \mathcal{S}^{\pm} $ are always adjoint with $ \mathcal{B}_{k}^{\mp} $. Then by taking the partial trace over the target, one obtains the reduced density of the quantum sensor:
\begin{equation}
      \hat{\rho}_{\rm S}\left(t\right)\approx\sum_{\eta_{k}\in\left\{\pm,0\right\}}\left(2\delta t\right)^{\Theta}C^{\eta_{N}\cdots\eta_{1}}\left(\mathcal{T}\prod_{k=1}^{N}\mathcal{S}^{\overline{\eta}_{k}}\right)\hat{\rho}_{\rm S},\label{reduced}
\end{equation}
which is completely determined by all types of quantum correlations $ C^{\eta_{N}\cdots\eta_{1}}=\Tr_{\mathrm{B}}\left(\mathcal{B}_{N}^{\eta_{N}}\cdots\mathcal{B}_{2}^{\eta_{2}}\mathcal{B}_{1}^{\eta_{1}}\hat{\rho}_{\mathrm{B}}\right)$ and $ \Theta$ defines the order number of the correlation  $ C^{\eta_{N}\cdots\eta_{1}}$. In general, direct measurement on the quantum sensor would involve all possible kinds of quantum correlations. 

To selectively extract arbitrary types of quantum correlations $ C^{\eta_{N}\cdots\eta_{1}} $ via the quantum sensor, we insert a general `quantum channel' (denoted by a super-operator $ \mathcal{P}_{k} $) before each short-time ($ \delta t $)-coupling evolution, as shown in Fig.~\ref{Fig1}b. Such a set of quantum channels can be realized by some unitary or non-unitary operations applied on the quantum sensor, such as rotation, measurement or polarization (see Supplementary Note 2). Then the final state of the quantum sensor turns into
\begin{equation}
    \hat{\rho'}_{\rm S}\left(t\right)\approx\sum_{\eta_{k}\in\left\{\pm,0\right\}}\left(2\delta t\right)^{\Theta}C^{\eta_{N}\cdots\eta_{1}}\left(\mathcal{T}\prod_{k=1}^{N}\mathcal{S}^{\overline{\eta}_{k}}\mathcal{P}_{k}\right)\hat{\rho}_{\rm S}.\label{state2}
\end{equation}
After measuring the observable $ \hat{O} $ on the final state of the quantum sensor, the obtained signal is
\begin{equation}
    S_{N}= \Tr_{\rm S}\left[\hat{O} \hat{\rho'}_{\rm S}\left(t\right)\right] \approx\sum_{\eta_{k}\in\left\{\pm,0\right\}}\delta t^{\Theta}A^{\overline{\eta}_{N}\cdots\overline{\eta}_{1}}C^{\eta_{N}\cdots\eta_{1}},\label{signal}
\end{equation}
where the coefficient is
\begin{equation}
    A^{\overline{\eta}_{N}\cdots\overline{\eta}_{1}}=2^{\Theta}\mathrm{Tr}_{\mathrm{S}}\left[\hat{O}\left(\mathcal{T}\prod_{k=1}^{N}\mathcal{S}^{\overline{\eta}_{k}}\mathcal{P}_{k}\right)\hat{\rho}_{\mathrm{S}}\right].\label{coefficient}
\end{equation}

In order to selectively extract arbitrary types of quantum correlations, we can design a set of quantum channels $ \left\{\mathcal{P}_{k}\right\} (k=1\cdots,N $) together with an observable $ \hat{O} $, so that only one term of equation~\eqref{signal} is reserved, e.g.,that associated with the correlation $ C^{\gamma_{N}\cdots\gamma_{1}} $. To achieve this goal, the channel sequence $ \left\{\mathcal{P}_{k}\right\}$ and observable $ \hat{O} $ have to make sure the coefficient $ A^{\overline{\gamma}_{N}\cdots\overline{\gamma}_{1}}\neq 0 $, while all other $ A^{\overline{\eta}_{N}\cdots\overline{\eta}_{1}} $ vanish for $ \eta_{N}\cdots\eta_{1}\neq\gamma_{N}\cdots\gamma_{1} $. Fig.~\ref{Fig1}b visualizes this idea. When passing through each $ \delta t $ slice, the quantum target has three evolution options: $ \mathcal{B}^{+}_{k} $, $ \mathcal{B}^{-}_{k} $ and $ \mathds{1} $. If the designed quantum channels $ \mathcal{P}_{k} $ are inserted between the neighbouring $ \delta t $ slices, only one connected path that leads to the desired correlation is reserved while other evolving paths of the quantum target are blocked. The extraction of arbitrary correlations in the case of more general coupling interaction ($\hat{V}=\sum_{\alpha=1}^{3}\hat{S}_{\alpha}\otimes\hat{B}_{\alpha}$) can be achieved by a generalized method, where the problem is reduced to solve the indefinite linear equations. It can be proved that it is always possible to find a solution because that the number of the linear equations is always less than the control elements (see Supplementary Note~2).

Next we will take the second-order correlation $ C^{+-} $ and the fourth-order correlation $ C^{+--+} $ as examples, where a spin-1/2 system is chosen as the quantum sensor coupled to a quantum target by pure dephasing spin model $ \hat{V}(t)=\frac{1}{2}\hat{\sigma}_{z}\otimes\hat{B}(t) $. Here $ \frac{1}{2}\hat{\sigma}_{z} $ is the $ z $ component of the sensor's spin operator, which corresponds to the super-operators of commutator $ \mathcal{S}_{z}^{-} $ and anti-commutator $ \mathcal{S}_{z}^{+} $. The initial state of the quantum sensor is $ \hat{\rho}_{\mathrm{S}}=\left(\mathds{1}+p\hat{\sigma}_{z}\right)/2 $ ($ p $ is the polarization of the sensor) and the final observable is $\hat{O}=\hat{\sigma}_{y}$. The quantum channels are constructed by synthesizing the spin rotations of the quantum sensor.

For measuring $ C^{+-} $, two channels $ \mathcal{P}_{1}^{\rm 2nd},\mathcal{P}_{2}^{\rm 2nd} $ are designed to reserve the coefficient $ A^{-+} $ while making other coefficients like $ A^{+-} $ and $ A^{--} $ vanish (see Methods). The scheme can be clearly illustrated by the channel diagram as shown in Fig.~\ref{Fig1}c. Since $\{\mathds{1},\hat{\sigma}_{x},\hat{\sigma}_{y},\hat{\sigma}_{z}\}$ constitutes a complete basis for the Liouville space of spin-$1/2$ system, the initial state $ \hat{\rho}_{\mathrm{S}} $ and observable $ \hat{O} $ can be represented by four-vectors. The super-operators like $ \mathcal{S}_{z}^{\eta_{k}} $ and $ \mathcal{P}_{k} $ are the $ 4\times4 $ matrices in this representation. $ \mathcal{S}_{z}^{\pm} $ are always adjoint with $ \mathcal{B}_{k}^{\mp} $. Consequently, the only non-vanishing coefficient $ A^{-+} $ associated with the correlation $ C^{+-} $ corresponds to the connected path (in solid line) that starts from $ \mathds{1} $ of the initial state and ends at $ \hat{\sigma}_{y} $ measurement of the final state (see Fig.~\ref{Fig1}c). As given in Methods, the coefficient $ A^{-+}=1 $, thus the measurement signals of $ \left<\hat{\sigma}_{y}\right> $ on the quantum sensor give the information of second-order correlation $ S_2 $ for $n =2$ in equation~\eqref{signal}].

The measurement of $ C^{+--+} $ can be realized by four quantum channels $ \mathcal{P}_{1}^{\rm 4nd},\mathcal{P}_{2}^{\rm 4nd},\mathcal{P}_{3}^{\rm 4nd},\mathcal{P}_{4}^{\rm 4nd} $ (see Methods). As shown in Fig.~\ref{Fig1}d, the non-vanishing coefficient $ A^{-++-} $ associated with $ C^{+--+} $ corresponds to the connected path (in solid line) that starts from $ \hat{\sigma}_{z} $ of the initial state and ends at $ \hat{\sigma}_{y} $ measurement of the final state. The other irrelevant paths are blocked because the coefficients related to them all vanish. Since $ A^{-++-}=p $ (see Methods), the final measurement signals of $ \left<\hat{\sigma}_{y}\right> $ on the quantum sensor give the information of fourth-order correlation $ S_{4} $ for $N=4$ in  equation~\eqref{signal}.

It is worth noting that some quantum correlations are inaccessible to the synthesized spin rotations of a spin-1/2 quantum sensor. For instance, the extraction of the third-order correlation $ C^{+-+} $ or any correlation like $ \mathcal{C}^{+-\cdots-+} $ with an odd number of commutators ($ \mathcal{B}_{k}^{-} $) requires the quantum channels that connect $ \mathds{1} $ with $ x/y/z $, which can't be synthesized by the rotations of a single spin-1/2 system. To measure them then, quantum channels beyond synthesized spin rotations or multi-spin quantum sensors are required.

\section*{Experimental demonstration}
The scheme above is experimentally demonstrated by using nuclear spins at room temperature on a Bruker Avance III 400 MHz nuclear magnetic resonance (NMR) spectrometer. The sample is carbon-13 labeled acetic acid (\textsuperscript{13}CH\textsubscript{3}COOH) dissolved in heavy water (D\textsubscript{2}O). The methyl group (-\textsuperscript{13}CH\textsubscript{3}) in acetic acid can be seen as a central spin system, where the \textsuperscript{13}C nucleus is the central spin as the sensor while three \textsuperscript{1}H nuclei constitute the quantum many-body target. The natural Hamiltonian of the system in doubly rotating frame is the coupling term $ \hat{H}_{\rm NMR} =\frac{\pi}{2}J_{\rm CH}\hat{\sigma}_{z}^{\rm C}\otimes\sum_{i=1}^{3}\hat{\sigma}_{i,z}^{\mathrm{H}} $ with the coupling constant $ J_{\rm CH} = 129.6 $ Hz, and directly generates the pure dephasing Hamiltonian $ \hat{V}=\frac{1}{2}\hat{\sigma}_{z}\otimes\hat{B}$ between the sensor and the target, where the target operator is $ \hat{B}=\frac{1}{2} J\sum_{i=1}^{3}\hat{\sigma}_{i,z}^{\mathrm{H}}$ with $ J=2\pi J_{\rm CH} $.

Figure~\ref{Fig2}a and Figure~\ref{Fig3}a show the experimental procedures for measuring  the second-order correlation $ C^{+-} $ and the fourth-order correlation $ C^{+--+} $, respectively. The system is initially prepared in a separable equilibrium state $ \hat{\rho}\left(0\right)=\hat{\rho}_{\mathrm{S}}\otimes\hat{\rho}_{\mathrm{B}} $ with $ \hat{\rho}_{\mathrm{S}}=\left(\mathds{1}+p_{\mathrm{C}}\hat{\sigma}_{z}^{\rm C}\right)/2 $ and $ \hat{\rho}_{\mathrm{B}}=\left(\mathds{1}+p_{\mathrm{H}}\sum_{i=1}^{3}\hat{\sigma}_{i,x}^{\rm H}\right)/8 $, where $\mathds{1}$ is the unit operator and $ p_{\mathrm{C}}, p_{\mathrm{H}} $ are the thermal polarizations ($ \sim 10^{-5} $) for the \textsuperscript{13}C and \textsuperscript{1}H spins, respectively.
A continuous radio frequency (RF) field on resonance along $ x $ axis is applied on the \textsuperscript{1}H spins to create the local Hamiltonian of the target: $ \hat{H}_{\mathrm{B}}=\pi\nu\sum_{i=1}^{3}\hat{\sigma}_{i,x}^{\rm H}$, where $ \nu\approx 24000 $ Hz is the nutation frequency of \textsuperscript{1}H spins. The quantum channels are constructed by the synthesis of $ \pi/2 $ pulses $\mathcal{R}_{\alpha}\left(\pi/2\right)$ with different phases $\alpha$. In experiments, they can be well realized by the phase cycling technology in NMR\cite{Bain1984}.
Finally, the signals are recorded by measuring the magnetizations of central spins (\textsuperscript{13}C) along $y$ axis, i.e., $ \left<\hat{\sigma}_{y}^C\right> $, from which arbitrary correlations can be extracted by different quantum circuits with suitable phase cycling schemes. 

Figure~\ref{Fig2}b presents the measured values of $ \left<\hat{\sigma}_{y}^C\right> $ versus the evolving time $ \tau_{21}=t_2-t_1 $ for $ \delta t=0.1~\&~0.5 $ ms (black scatters), from which the second-order correlation $ C^{+-} $ [equation~\eqref{C2}] is extracted. The target signals $ \delta t^{2}C^{+-}(t_2,t_1) $ (solid red lines) and the numerical results of $ S_{2} $ [equation~\eqref{methods2th}] (black dashed lines) are also presented. As theoretically expected, the measured signals show oscillatory behavior with $\tau_{21}$. The relative deviation (defined in Methods) between the measured signals and the target signals is $\Delta=32.3\% $ for $ \delta t=0.1 $ ms and $ \Delta=18.8\% $ for $ \delta t=0.5 $ ms. 
We also plot the dependence of $ \left<\hat{\sigma}_{y}^C\right> $ on $ J\delta t $ with a fixed interval $ \tau_{21}=10 $ $\mu$s in Fig.~\ref{Fig2}c. From this we find that with the increase of $ \delta t$, the experimental data show increasing deviation from the target signals (red solid line), but agree well with the numerical simulation (dashed line). The deviation between the numerical simulation and the target signals comes from the theoretical approximation due to the finite value of $ \delta t $. Moreover, we fit the measured signals $ \left<\hat{\sigma}_{y}^C\right> $ with a power law of $ \left(J\delta t\right)^{k} $ and find $ k\approx 1.856 $, which is close to the ideal value $ k=2 $. 
A little smaller $k$ is mainly caused by the higher-order contribution in $ S_{2} $ due to the finite value of $ \delta t $ [see equation~\eqref{methods2th}]. As expected, we can see from Fig.~\ref{Fig2}c that the deviation is negligible when $ J\delta t\to 0 $ while it becomes remarkable when $ J\delta t\to 1 $. Meanwhile, we can also find from the inset that the fitting increasingly deviates from the linearity and the fitting parameter $k$ becomes unstable due to the lower signal-to-noise ratios (SNRs) when $ J\delta t\to 0 $. Therefore, there exists a trade-off between the theoretical approximation and the SNR of the measured signals resulting from $ \delta t $. The optimal value of $ \delta t $ can be obtained with the aid of the numerical simulations (see Methods and Supplementary Note 4), and $ \delta t = 0.5  $ ms is a relatively suitable time to extract the second-order correlations. Besides the errors from the theoretical approximation, other error mechanisms leading to the deviation between the experimental data and the numerical simulation include the control imperfection of the $\pi/2$-pulses, the evolution error of the quantum target caused by the RF inhomogeneity and the readout error from the final measurement. We numerically analyze the contributions of these errors, where the error from the $\pi/2$-pulse imperfection is very small in the measurement of the second-order correlations (see Supplementary Table~3).

It is more difficult to measure the fourth-order correlations since the target signals will be much weaker than those of the second-order correlations. As shown in equation~\eqref{methods4th}, the signal related to $ C^{+--+} $ is proportional to $\delta t ^4$, which leads to higher requirements for the experiments, including the higher sensitivity of the quantum sensor and the higher control accuracy of the quantum channels. While the $ \pi/2 $-pulse imperfection (about $ 2\%\sim3\% $ relative error) is negligible in the measurement of the second-order correlations, it will bring a considerable impact on the measurement of the fourth-order correlations. As analyzed in Methods, the non-ideal channel $ \mathcal{P}_{3}^{\rm 4nd} $ will result in a severe leakage of lower-order correlations to the measurement of the four-order correlations and overwhelm the desired signal $ C^{+--+} $, which brings a challenge in measurements. Hence the scheme as shown in Fig.~\ref{Fig1}d is practically infeasible in experiments. To overcome this problem, we design an error-resilient channel by repeating the non-ideal $ \mathcal{P}_{3}^{\rm 4th} $ for $ n $ times to weaken the unwanted low-order correlations (see Fig.~\ref{Fig3}a). As demonstrated in equation~\eqref{methods4E}, the robust channel $ \left(\mathcal{P}_{3}^{\rm 4th}\right)^n $ exponentially approaches to the ideal channel with an error of order $ \delta\theta^{n} $ as $ n $ increases. Therefore, the unwanted lower-order signals ($ \delta t^{2}C^{+00+} $) can be effectively suppressed. 

\begin{figure}
    \centering
    \includegraphics[scale=0.95]{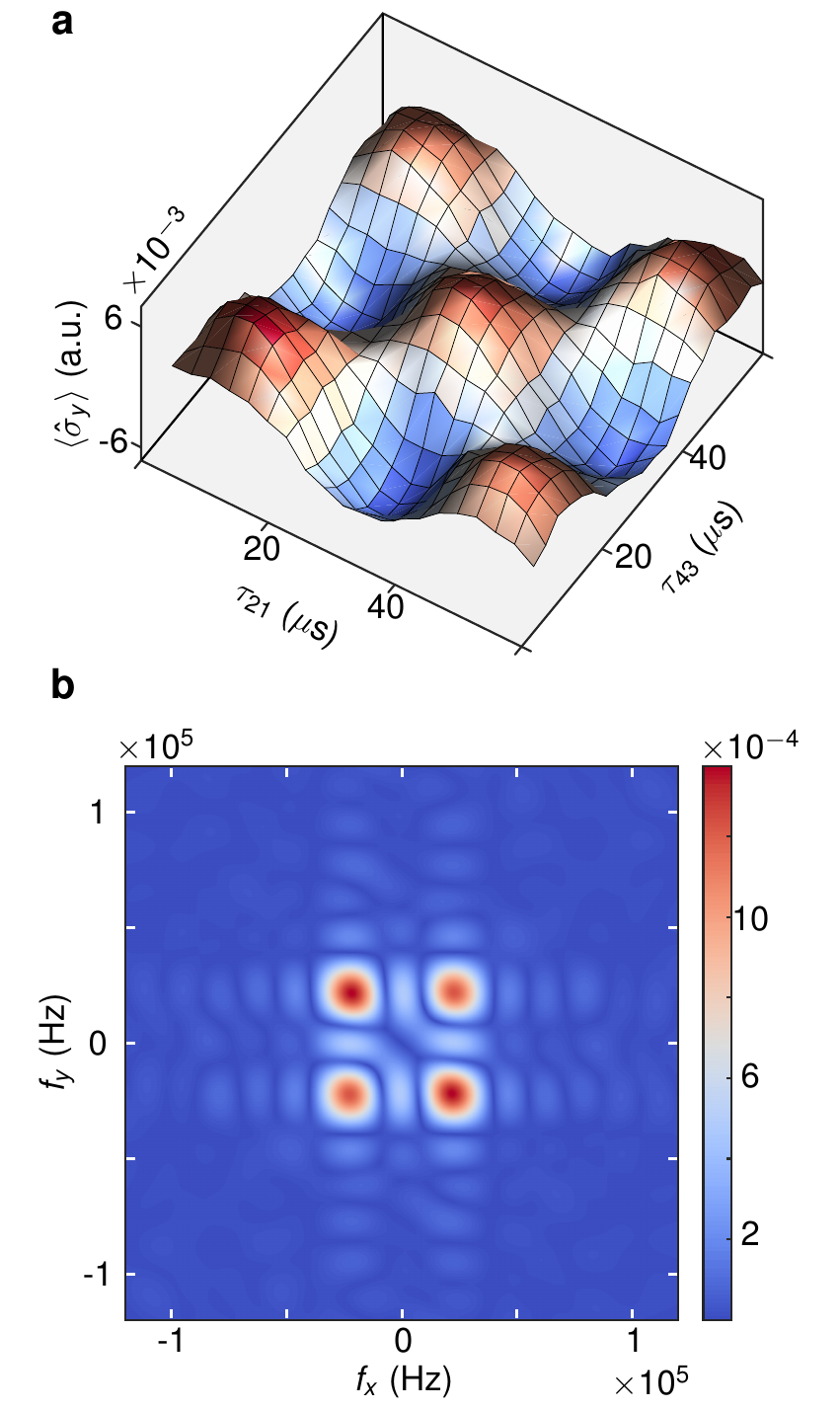}
    \caption{\textbf{Experimental measurement of the 2D structure of the fourth-order correlations $ C^{+--+} $.} \textbf{a}, Measured signals $ S_{4}(\tau_{21}, \tau_{43}) $ with the robust quantum channel. \textbf{b}, Spectral density of the fourth-order correlation $ C^{+--+}(f_x,f_y) $, which is obtained from the 2D Fourier transform of $ S_{4}(\tau_{21}, \tau_{43}) $.}\label{Fig4}
\end{figure}
Figure~\ref{Fig3}b presents the measurement values of $ \left<\hat{\sigma}_{y}^C\right> $ versus the evolving time $ \tau_{21}=t_2-t_1 $, including both the cases of $ n=1 $ and $ n=3 $. Here $ \tau_{32}=t_3-t_2 $ and $ \tau_{43}=t_4-t_3 $ are fixed at 10 $\mu$s, and $ \delta t=0.5 $ ms. The target signal $ \delta t^{4}p_{\rm C}C^{+--+}(t_4,t_3,t_1,t_1) $ (solid red line) and the numerical results of $ S^{\rm E}_{4} $ [equation~\eqref{methods4E}] with (i.e., n = 3, black dashed line) and without (i.e., n=1, blue dashed line) the robust channels for the pulse imperfections, are also presented. For the robust channels ($ n=3 $), the measured signals (black scatters) agree well with the target signals, and the relative deviation between them is $ \Delta=23.5\% $. By contrast, for the non-robust channels ($ n=1 $), the measured signals (blue scatters) show serious deviation from the target signals ($ \Delta=112.2\% $), which makes the data almost untrusted for measuring $ C^{+--+} $. For the robust channels, we also plot the dependence of $ \left<\hat{\sigma}_{y}^C\right> $ on $ J\delta t $ with a fixed interval $ \tau_{21}=10 $ $\mu$s in Fig.~\ref{Fig3}c. Similar to the results of measuring $ C^{+-} $, the experimental data (black scatters) are in good agreement with the numerical simulations (dashed black line), but gradually deviate the target signals (red solid line) when $ J\delta t\to 1 $. As shown in the inset, the power law fitting of the measured signals $ \left<\hat{\sigma}_{y}^C\right>\propto\left(J\delta t\right)^{k} $ gives the exponent $ k=3.492 $, where the deviation from the ideal prediction $ k=4 $ is also mainly caused by the higher-order contributions to $ S_{4} $ due to the finite value of $ \delta t $. Besides the approximation errors from $ \delta t $, the main experimental errors of measuring the fourth-order correlations by using the robust channels are the evolution error of the quantum target caused by RF inhomogeneity and the readout error, while the control error caused by $\pi/2$-pulse imperfection is almost negligible (see Supplementary Table~4). We also analyze the trade-off between the theoretical approximation and the SNR of the measured signals resulting from $ \delta t $, and $ \delta t = 0.5  $ ms corresponds to relatively low errors (see Methods). 

To achieve the complete fourth-order quantum correlation $ C^{+--+}\left(t_{4},t_{3},t_{2},t_{1}\right) $, we measure the signals of $ S_{4} $ versus the evolving time $ \tau_{21} $ and $ \tau_{43} $ (note that $ \mathcal{C}^{+--+} $ doesn't depend on $ \tau_{32} $) using the robust quantum channels (see Fig.~\ref{Fig4}a). The measured signals show two-dimensional oscillatory behavior along with $\tau_{21}$ and $ \tau_{43} $, and the spectral density are obtained from the 2D Fourier transform of $ S_{4} $ as shown in Fig.~\ref{Fig4}b.

\section*{Conclusion}

We demonstrate selective measurement of arbitrary types and orders of quantum correlations via the synthesized quantum channels with a quantum sensor. The correlation-selection approach based on synthesized quantum channels is more universal than the previously proposed weak measurement scheme in that the former is also applicable to ensemble systems. We successfully extract the second- and fourth-order correlations in a system of nuclear spins with a spin-1/2 sensor. The experiment can be generalized to the quantum sensors of higher or multiple spins, as well as bosonic or fermionic systems. Compared with the conventional nonlinear spectroscopy, this scheme exponentially broadens the accessible correlations. Higher order correlations provide new important information about quantum many-body systems that is not available from conventional nonlinear spectroscopy\cite{SchweiglerNature2017}. Our work offers a new approach to understanding quantum many-body physics (e.g., many-body localization\cite{ChoiNature2017,Fan2017} and quantum thermalization\cite{KaufmanScience2016,Swingle2018}), to examining the quantum foundation (e.g., quantum nonlocality\cite{Popescu2014}), and to providing key information for quantum technologies (e.g., the characterization of quantum computers and quantum simulators, the optimization of quantum control and the improvement of quantum sensing\cite{DegenNature2019}).


\section*{Methods}

\subsection*{Measuring $ C^{+-} $.}
According to equation~\eqref{signal}, the second-order correlation $ C^{+-} $ corresponds to the target signal
\begin{equation}
    S_{2}^{\rm target}=\delta t^{2}A^{-+}C^{+-},
\end{equation}
with the coefficient 
\begin{equation}
    A^{-+}=2^{2}\Tr_{\mathrm{S}}\left(\hat{O}\mathcal{S}_{z}^{-}\mathcal{P}_{2}^{\rm 2nd}\mathcal{S}_{z}^{+}\mathcal{P}_{1}^{\rm 2nd}\hat{\rho}_{\mathrm{S}}\right).
\end{equation}
To selectively extract the second-order correlation $ C^{+-} $ under the current experimental setup, we need to make $ A^{-+} $ become the only non-vanishing coefficient. Then with the help of the channel diagram in Fig.~\ref{Fig1}c, the quantum channels $ \mathcal{P}_{1,2}^{\rm 2nd} $ are designed to be
\begin{equation}
    \begin{aligned}
         \mathcal{P}_{1}^{\rm 2nd}&=\left[\mathcal{R}_{y}\left(\pi/2\right)+\mathcal{R}_{-y}\left(\pi/2\right)\right]/2,\\
         \mathcal{P}_{2}^{\rm 2nd}&=\mathcal{R}_{y}\left(\pi/2\right),
    \end{aligned}\label{CHC2}
\end{equation}
where $ \mathcal{R}_{\alpha}\left(\pi/2\right) $ is the $ \pi/2 $ spin rotation along $ \alpha $-axis and $ \mathcal{P}_{1}^{\rm 2nd} $ is realized by the combination of two $ \pi/2 $ rotations. 

In the representation of $ \{\mathds{1},\hat{\sigma}_{x},\hat{\sigma}_{y},\hat{\sigma}_{z}\} $, the matrix forms of $ \mathcal{S}_{z}^{\pm} $ are 
\begin{equation}
    2\mathcal{S}_{z}^{+}=\begin{pmatrix}0 & 0 & 0 & 1\\
        0 & 0 & 0 & 0\\
        0 & 0 & 0 & 0\\
        1 & 0 & 0 & 0
        \end{pmatrix},\qquad
        2\mathcal{S}_{z}^{-}=\begin{pmatrix}0 & 0 & 0 & 0\\
        0 & 0 & -1 & 0\\
        0 & 1 & 0 & 0\\
        0 & 0 & 0 & 0
        \end{pmatrix},
\end{equation}
and the matrix representation of $ \mathcal{R}_{a}\left(\theta\right) $ is
\begin{equation}
    \mathcal{R}_{\alpha}\left(\theta\right)=
    \begin{pmatrix}
        1 & 0\\
        0 & \mathbf{R}\left(\theta,\mathbf{n}_{\alpha}\right)
    \end{pmatrix}.
\end{equation}
Here $ \mathbf{R}\left(\theta,\mathbf{n}_{\alpha}\right) $ is the 3D rotation along the axis $ \mathbf{n}_{\alpha} $. The matrix elements ($ R_{ij} $) of $\mathbf{R}\left(\theta,\mathbf{n}\right) $ for $ \mathbf{n}=\left(n_{x},n_{y},n_{z}\right) $ are 
\begin{equation}
    R_{ij}\left(\theta,\mathbf{n}\right)=\delta_{ij}\cos\theta+n_{i}n_{j}(1-\cos\theta)+\sin\theta\sum_{l=1}^{3}\epsilon_{ijk}n_{k},
\end{equation}
where $ i,j,k\in \left\{x,y,z\right\} $, and $ \epsilon_{ijk} $ is the Levi-Civita symbol. Since a spin-1/2 quantum sensor has the vector form of density matrix $ \vec{\rho}=\left(1,p_x,p_y,p_{z}\right)/2 $, whose measurement result of $ \hat{\sigma}_{i} $ is $ \Tr\left(\hat{\sigma}_{i}\rho\right)=2p_{i} $. Then with $ \hat{\rho}_{\mathrm{S}}=(\mathds{1}+p\hat{\sigma}_{z})/2 $ and $ \hat{O}=\hat{\sigma}_{y} $, the coefficient is calculated to be $ A^{-+}=1 $, while other coefficients such as $ A^{0\pm} $ and $ A^{\pm 0} $ vanish. Hence the measurement signal is
\begin{equation}
    S_{2}=\delta t^{2}C^{+-}(t_{2},t_{1})+O\left(\delta t^{4}\right).\label{methods2th}
\end{equation}

Considering the $ \pi/2 $-pulse imperfection with the angle error ($ \pi/2 \to \pi/2+\delta\theta $), the measurement signal becomes
\begin{equation}
    \begin{aligned}
    S^{\rm E}_{2}&=\cos\left(\delta\theta\right)\delta t^{2}C^{+-}(t_{2},t_{1})+O\left(\delta t^{4}\right) \\
    &\approx \left(1-\delta\theta^2\right)\delta t^{2} C^{+-}(t_{2},t_{1})+O\left(\delta t^{4}\right).
    \end{aligned}\label{2thE}
\end{equation}
Then the pulse imperfection consequently introduces an error of order $\delta\theta^{2}$ to the amplitude of the measured signal, which is usually small. 

For the initial state of the quantum target in experiment $ \hat{\rho}_{\mathrm{B}}=\left(\mathds{1}+p_{\mathrm{H}}\sum_{i=1}^{3}\hat{\sigma}_{i,x}^{\rm H}\right)/8 $, the target operator $ \hat{B}=\frac{1}{2} J\sum_{i=1}^{3}\hat{\sigma}_{i,z}^{\mathrm{H}} $ and the local Hamiltonian $ \hat{H}_{\mathrm{B}}=\pi\nu\sum_{i=1}^{3}\hat{\sigma}_{i,x}^{\mathrm{H}} $, the specific form of $ C^{+-} $ is
\begin{equation}
    \begin{aligned}
         C^{+-}&=\Tr\left[\mathcal{B}^{+}(t_{2})\mathcal{B}^{-}(t_{1})\hat{\rho}_{\mathrm{B}}\right]\\
         &=\frac{3}{4}J^{2} p_{\rm H}\sin\left[2\pi\nu\left(t_{2}-t_{1}\right)\right],
    \end{aligned}\label{C2}
\end{equation}
Obviously, the target signal $ \delta t^{2}C^{+-} $ quadratically depends on $ J\delta t $.

\subsection*{Measuring $C^{+--+}$.}
According to equation~\eqref{signal}, the fourth-order correlation $ C^{+--+} $ corresponds to the target signal
\begin{equation}
    S_{4}^{\rm target}=\delta t^{4}A^{-++-}C^{+--+},
\end{equation}
with the coefficient 
\begin{equation}
    A^{-++-}=2^{4}\Tr_{\mathrm{S}}\left(\hat{O}\mathcal{S}_{z}^{-}\mathcal{P}_{4}^{\rm 4th}\mathcal{S}_{z}^{+}\mathcal{P}_{3}^{\rm 4th}\mathcal{S}_{z}^{+}\mathcal{P}_{2}^{\rm 4th}\mathcal{S}_{z}^{-}\mathcal{P}_{1}^{\rm 4th}\hat{\rho}_{\mathrm{S}}\right).
\end{equation}
To selectively extract the fourth-order correlation $ C^{+--+} $ under the current experimental setup, we need to make $ A^{-++-} $ become the only non-vanishing coefficient. Then with the help of the channel diagram in Fig.~\ref{Fig1}d, the quantum channels $ \mathcal{P}_{1,2,3,4}^{\rm 4th} $ are designed to be
\begin{equation}
    \begin{aligned}
    \mathcal{P}_{1}^{\rm 4th}=\mathcal{P}_{4}^{\rm 4th}&=\left[\mathcal{R}_{y}(\pi/2)-\mathcal{R}_{-y}(\pi/2)\right]/2,\\
    \mathcal{P}_{2}^{\rm 4th}&=\left[\mathcal{R}_{x}(\pi/2)-\mathcal{R}_{-x}(\pi/2)\right]/2,\\
    \mathcal{P}_{3}^{\rm 4th}&=\left[\mathcal{R}_{x}(\pi/2)+\mathcal{R}_{-x}(\pi/2)\right]/2.
    \end{aligned}
\label{eq:dissipation channel}
\end{equation}
With $ \hat{\rho}_{\mathrm{S}}=(\mathds{1}+p_{\rm C}\hat{\sigma}_{z})/2 $ and $ \hat{O}=\hat{\sigma}_{y} $, the only non-vanishing coefficient is calculated to be $ A^{-++-}=p $. Then the measurement signal is
\begin{equation}
    S_{4}=\delta t^{4}p_{\rm C} C^{+--+}\left(t_{4},t_{3},t_{2},t_{1}\right)+O\left(\delta t^{6}\right).\label{methods4th}
\end{equation}

Similar to the case of measuring $ C^{+-} $, the non-ideal quantum channels caused by $ \pi/2 $-pulse imperfection ($\pi/2 \to \pi/2+\delta\theta$) will introduce an overall coefficient $ \cos\left(\delta\theta\right) $ to $ \mathcal{P}_{1,2,4}^{\rm 4th} $, i.e., $ \mathcal{P}_{1,2,4}^{\rm 4th} \to \cos\left(\delta\theta\right)\mathcal{P}_{1,2,4}^{\rm 4th} $. However, the error mechanism of $ \mathcal{P}_{3}^{\rm 4th} $ is totally different because two extra matrix elements proportional to $ \sin\left(\delta\theta\right) $ are introduced:
\begin{equation}
    \mathcal{P}_{3}^{\rm 4th}\to
        \begin{pmatrix}1 & 0 & 0 & 0\\
            0 & 1 & 0 & 0\\
            0 & 0 & -\sin\left(\delta\theta\right) & 0\\
            0 & 0 & 0 & -\sin\left(\delta\theta\right)
        \end{pmatrix}.
\end{equation}
In this situation, an extra path related to lower-order correlation $ C^{+00+} $ will be connected and get mixed with $ C^{+--+} $ in the final measurement signals, i.e.,
\begin{equation}
    \begin{aligned}
        S_{4}^{\rm E}&= p_{\rm C}\cos\left(\delta\theta\right)^{3}\delta t^{4}C^{+--+}\left(t_{4},t_{3},t_{2},t_{1}\right)\\
        &\qquad+p_{\rm C}\sin\left(\delta\theta\right)\delta t^{2}C^{+00+}\left(t_{4},t_{1}\right)+O\left(\delta t^{6}\right).
    \end{aligned}
\end{equation}
Consequently, the non-ideal channels $ \mathcal{P}_{1,2,4}^{\rm 4th} $ only bring a total scaling factor $ \cos\left(\delta\theta\right)^{3}\approx 1 $ to the fourth-order correlation $ C^{+--+} $, which can be easily corrected by the error calibration. But the imperfection of $ \mathcal{P}_{3}^{\rm 4th} $ has greater impact on measuring $ C^{+--+} $ for two reasons: On the one hand, the non-ideal $ \mathcal{P}_{3}^{\rm 4th} $ introduces the lower-order correlation signals of $ \delta t^2 $ scale, which are much larger than the signals of $ C^{+--+} $ ($ \sim \delta t^4 $). On the other hand, the error order of $ \sin\left(\delta\theta\right)\approx\delta\theta $ is also lower than $ \cos\left(\delta\theta\right)\approx\left(1-\delta\theta^{2}/2\right) $.

To deal with the major error source, an error-resilient quantum channel is constructed by repeating the non-ideal $ \mathcal{P}_{3}^{\rm 4th} $ for $ n $ times, whose matrix form is
\begin{equation}
    \left(\mathcal{P}_{3}^{\rm 4th}\right)^n=
    \begin{pmatrix}
        1 & 0 & 0 & 0\\
        0 & 1 & 0 & 0\\
        0 & 0 & -\sin\left(\delta\theta\right)^n & 0\\
        0 & 0 & 0 & -\sin\left(\delta\theta\right)^n
    \end{pmatrix}.
\end{equation}
As shown in Figure~\ref{Fig3}, for $ \hat{\rho}_{\mathrm{S}}=(1+p_{\rm C}\hat{\sigma}_{z})/2 $ and $ \hat{O}=\hat{\sigma}_{y} $, the non-vanishing coefficients are calculated to be $ A^{-++-}\approx p_{\rm C}\left(1-\delta\theta^{2}/2\right)^{3} $ and $ A^{-00-}\approx p_{\rm C}\left(-\delta\theta\right)^{n}$. So the measured signal using the robust channel becomes 
\begin{equation}
    \begin{aligned}
        S_{4}^{\rm E}&\approx  p_{\rm C}\left(1-\delta\theta^{2}/2\right)^{3}\delta t^{4}C^{+--+}\left(t_{4},t_{3},t_{2},t_{1}\right)\\
        &\qquad+p_{\rm C}(-\delta\theta)^{n}\delta t^{2}C^{+00+}\left(t_{4},t_{1}\right)+O\left(\delta t^{6}\right),
    \end{aligned}\label{methods4E}
\end{equation}
which reduces to equation~\eqref{methods4th} without $ \pi/2 $-pulse error ($ \delta\theta=0 $). Since the robust channel $ \left(\mathcal{P}_{3}^{\rm 4th}\right)^n $ exponentially approaches to the ideal channel with an error of order $ \delta\theta^{n} $ as $ n $ increases, the unwanted low-order contribution ($ \delta t^{2}C^{+00+} $) in the measured signals can be effectively suppressed. 

With the same experimental setup as measuring $ C^{+-} $, the specific form of $ C^{+--+} $ is
\begin{equation}
    \begin{aligned}
         C^{+--+}&=\Tr\left[\mathcal{B}^{+}(t_{4})\mathcal{B}^{-}(t_{3})\mathcal{B}^{-}(t_{2})\mathcal{B}^{+}(t_{1})\hat{\rho}_{\mathrm{B}}\right]\\
         &=\frac{3}{16}J^{4}\sin\left[2\pi\nu\left(t_{2}-t_{1}\right)\right]\sin\left[2\pi\nu\left(t_{4}-t_{3}\right)\right],
    \end{aligned}\label{C4}
\end{equation}
which means the target signal ($ \delta t^{4}C^{+--+} $) quartically depends on $ J\delta t $. Moreover, the specific form of the lower-order error term $ C^{+00+} $ is
\begin{equation}
    \begin{aligned}
         C^{+00+}&=\Tr\left[\mathcal{B}^{+}(t_{4})\mathcal{B}^{-}(t_{1})\hat{\rho}_{\mathrm{B}}\right]\\
         &=\frac{3}{4} J^{2}\cos\left[2\pi\nu\left(t_{4}-t_{1}\right)\right].
    \end{aligned}
\end{equation}
Therefore, to eliminate the major error source from the component of $C^{+00+}$ in equation~\eqref{methods4E}, we need to ensure
\begin{equation}
    \frac{1}{4}\left(1-\delta\theta^{2}/2\right)^{3}\left(J\delta t\right)^{2}\gg \left|\delta\theta\right|^{n}.\label{condition}
\end{equation}
For the angle error $ \delta\theta=0.04 $ rad from experimental error calibration and $ \delta t= 0.5 $ ms, we can deduce that $ n>1 $ is required to eliminate the lower-order signals. 

\subsection*{Error analysis.}
To quantify the deviation between the experimental data array $ \vec{x}=\left\{x(\tau_i)\right\} $ and the theoretical data array $ \vec{y}=\left\{y(\tau_i)\right\} $ ($ \tau_i $ is the sampling time point), we use the absolute error $ \vec{E} $ and relative error $ \Delta $ defined as:
\begin{equation}
  \vec{E}=\left|\vec{x}-\vec{y}\right|,\quad \Delta=\norm{\vec{E}}/\norm{\vec{y}}, \label{errdef}
\end{equation}
where $ \norm{\vec{x}}=\sqrt{\sum_i |x(\tau_i)|^2} $ is the Euclidean norm. The theoretical approximation error of finite $ \delta t $ can be measured by 
\begin{equation}
  \vec{E}^{\rm th}=\left|\vec{S}^{\rm sim}-\vec{S}^{\rm target}\right|,
\end{equation}
where $ \vec{S}^{\rm target} $ is the target signals, and $ \vec{S}^{\rm sim} $ is the simulated signals of $ S_{2} $ or $ S_{4} $. Besides that, other errors in experiments include the control error caused by $\pi/2$-pulse imperfection ($ \vec{E}^{\pi/2} $), the evolution error of the quantum target caused by RF inhomogeneity ($ \vec{E}^{\rm evo} $), and the experimental readout error ($ \vec{E}^{\rm r} $). After characterizing the deviation of $\pi/2$-pulse, the decay rate of the oscillatory signals, as well as the strength of spectral ground noise, these imperfect signals with only one type error can be numerically simulated, which are defined as $ \vec{S}^{\pi/2}, \vec{S}^{\rm evo} $ and $ \vec{S}^{\rm r} $ respectively. Then these error contributions can be investigated separately by calculating the deviations between these imperfect signals and the ideal signals $ \vec{S}^{\rm sim} $, i.e.,
\begin{equation*}
    \vec{E}^{\pi/2}=\left|\vec{S}^{\pi/2}-\vec{S}^{\rm sim}\right|,
    \vec{E}^{\rm evo}=\left|\vec{S}^{\rm evo}-\vec{S}^{\rm sim}\right|,
    \vec{E}^{\rm r}=\left|\vec{S}^{\rm r}-\vec{S}^{\rm sim}\right|.
\end{equation*}
Therefore, their relative errors, i.e., $ \Delta^{\pi/2},\Delta^{\rm evo} $ and $ \Delta^{\rm r} $, are presented in Supplementary Table 3 and 4 according to the definition~\eqref{errdef}.

\subsection*{Optimal coupling-evolution time.}
The scheme in experiments requires a relatively small $ \delta t $ for a high enough theoretical approximation. However, as the $\Theta$-order-correlation signals are proportional to $\delta t^{\Theta}$, smaller $ \delta t $ will lead to lower SNRs in practical measurement. Therefore, a trade-off between the theoretical approximation and the SNR of the measured signals can be described by the total relative error of the target correlation signals (see Supplementary Note 4):
\begin{equation*}
    \begin{aligned}
        \Delta^{\rm tot}\left(\delta t\right)&=\frac{\norm{\vec{E}^{\rm th}\left(\delta t\right)}+\norm{\vec{E}^{\rm \pi/2}\left(\delta t\right)}+ \norm{\vec{E}^{\rm evo}\left(\delta t\right)}+\norm{\vec{E}^{\rm r}}}{\norm{\vec{S}^{\rm target}\left(\delta t\right)}}\\
        &\approx \frac{\delta t^{\Theta+2}\norm{\vec{C}^{\eta_{N+2}\cdots\eta_{1}}}+\norm{\vec{E}^{\rm r}}}{\delta t^{\Theta}\norm{A^{\overline{\eta}_{N}\cdots\overline{\eta}_{1}}\vec{C}^{\eta_{N}\cdots\eta_{1}}}}+\Delta^{\pi/2}\left(\delta\theta\right)+\Delta^{\rm evo}
        \label{totE}
    \end{aligned} 
\end{equation*}
Here $ \vec{S}^{\rm target}\left(\delta t\right)=\delta t^{\Theta}A^{\overline{\eta}_{N}\cdots\overline{\eta}_{1}}\vec{C}^{\eta_{N}\cdots\eta_{1}} $ is the target signals of the desired correlations. $ \Delta^{\pi/2}\left(\delta\theta\right) $ is the relative error caused by the $ \pi/2 $-pulse imperfection ($\pi/2 \to \pi/2+\delta\theta$), which is derived from equations~\eqref{2thE} and \eqref{methods4E}:
\begin{equation}
    \Delta^{\pi/2}\left(\delta\theta\right) \approx
    \begin{cases}
    \delta\theta^2,\qquad\qquad\qquad\qquad~~~ {\rm for~~}C^{+-}, \\
    p_{\rm C}\left[1-\left(1-\delta\theta^{2}/2\right)^{3}\right],\quad {\rm for~~}C^{+--+}. 
    \end{cases}
\end{equation}
Note that the lower-order leakage in equation~\eqref{methods4E} is greatly suppressed when equation~\eqref{condition} is satisfied. $ \Delta^{\rm evo} $ is the relative error caused by RF inhomogeneity, i.e.,
\begin{equation}
    \Delta^{\rm evo}=\frac{\norm{\left(1-\ee^{-k\vec{\tau}}\right)\cdot\vec{C}^{\eta_{N}\cdots\eta_{1}}}}{\norm{\vec{C}^{\eta_{N}\cdots\eta_{1}}}}.
\end{equation}
Here $ k=2.76\times 10^{3} $ denotes the decay rate of the free evolution of the quantum target and $ \vec{\tau} $ is the sampling time list. $ \vec{E}^{\rm r} $ is determined by the ground noise of spectra and totally independent of $ \delta t $. Therefore, the $ \pi/2 $-pulse imperfection and rf inhomogeneity together contribute a constant relative error. Then, by taking $\partial \Delta^{\rm tot}_{\Theta}\left(\delta t\right) / \partial\Delta t=0$, the optimal evolution time is obtained:
\begin{equation}
    \delta t_{\rm opt}\approx\left(\frac{2\norm{\vec{C}^{\eta_{N+2}\cdots\eta_{1}}}}{\Theta\norm{\vec{E}^{\rm r}}}\right)^{1/(\Theta+2)}.
\end{equation}
With the experimental estimations of the parameters of the error sources (see Supplementary Note 3), all of these error sources can be simulated individually. Supplementary Figure~4 presents the simulated total relative error $ \Delta^{\rm tot}\left(\delta t\right) $ versus $ \delta t $ of measuring $ C^{+-} $ and $ C^{+--+} $. Then, the optimal $ \delta t $ with the smallest relative error can be obtained, i.e., $ \delta t_{\rm opt}=0.35 $ ms and 0.38 ms for  $ C^{+-} $ and $ C^{+--+} $ respectively. The coupling time $ \delta t=0.5 $ ms used in experiments also corresponds to relatively low errors.

\section*{Acknowledgements}
This work is supported by the National Key R \& D Program of China (Grants No. 2018YFA0306600 and 2016YFA0301203), the National Science Foundation of China (Grants No. 11822502, 11974125 and 11927811), Anhui Initiative in Quantum Information Technologies (Grant No. AHY050000), and Hong Kong Research Grants Council-General Research Fund Project 14300119

\section*{Author contributions}
R. B. L. and X. P. conceived the project. R.B.L., P. W. and Z. W. formulated the theoretical framework. X. P., Z. W. and Y. L. designed the experiment. Z. W., Y. L. and T. W. performed the measurements and analyzed the data. R. L. and Y. C. assisted with the experiment. X. P. and J. D. supervised the experiment. P. W., Z. W., R.B.L. and X.P. wrote the manuscript. All authors contributed to analyzing the data, discussing the results and commented on the writing.
\section*{Competing interests}
The authors declare no competing interests.

\end{document}


\title{Supplementary Information-Detection of arbitrary quantum correlations via synthesized quantum channels}

\begin{abstract}
The supplementary information contains some theoretical details about the construction methods of arbitrary quantum channels, the analyses of the main experimental error sources and their contributions, as well as some additional experimental data and details.
\end{abstract}

\author{Ze Wu}%
\thanks{These authors contribute equally}
\affiliation{
CAS Key Laboratory of Microscale Magnetic Resonance and School of Physical Sciences, University of Science and Technology of China, Hefei 230026, China}
\affiliation{
CAS Center for Excellence in Quantum Information and Quantum Physics, University of Science and Technology of China, Hefei 230026, China}

\author{Ping Wang}%
\thanks{These authors contribute equally}
\affiliation{
College of Education for the future, Beijing Normal University at Zhuhai (BNU Zhuhai), Zhuhai, China}
\affiliation{
Department of Physics, The Chinese University of Hong Kong, Shatin, New Territories, Hong Kong, China}

\author{Tianyun Wang}
\affiliation{
CAS Key Laboratory of Microscale Magnetic Resonance and School of Physical Sciences, University of Science and Technology of China, Hefei 230026, China}
\affiliation{
CAS Center for Excellence in Quantum Information and Quantum Physics, University of Science and Technology of China, Hefei 230026, China}

\author{Yuchen Li}
\affiliation{
CAS Key Laboratory of Microscale Magnetic Resonance and School of Physical Sciences, University of Science and Technology of China, Hefei 230026, China}
\affiliation{
CAS Center for Excellence in Quantum Information and Quantum Physics, University of Science and Technology of China, Hefei 230026, China}

\author{Ran Liu}
\affiliation{
CAS Key Laboratory of Microscale Magnetic Resonance and School of Physical Sciences, University of Science and Technology of China, Hefei 230026, China}
\affiliation{
CAS Center for Excellence in Quantum Information and Quantum Physics, University of Science and Technology of China, Hefei 230026, China}

\author{Yuquan Chen}
\affiliation{
CAS Key Laboratory of Microscale Magnetic Resonance and School of Physical Sciences, University of Science and Technology of China, Hefei 230026, China}
\affiliation{
CAS Center for Excellence in Quantum Information and Quantum Physics, University of Science and Technology of China, Hefei 230026, China}

\author{Xinhua Peng}%
\email{xhpeng@ustc.edu.cn}
\affiliation{
CAS Key Laboratory of Microscale Magnetic Resonance and School of Physical Sciences, University of Science and Technology of China, Hefei 230026, China}
\affiliation{
CAS Center for Excellence in Quantum Information and Quantum Physics, University of Science and Technology of China, Hefei 230026, China}

\author{Ren-Bao Liu}%
\email{rbliu@cuhk.edu.hk}
\affiliation{
Department of Physics, The Chinese University of Hong Kong, Shatin, New Territories, Hong Kong, China}
\affiliation{
Centre for Quantum Coherence, The Chinese University of Hong Kong, Shatin, New Territories, Hong Kong, China}
\affiliation{
The Hong Kong Institute of Quantum Information Science and Technology, The Chinese University of Hong Kong, Shatin, New Territories, Hong Kong, China}

\author{Jiangfeng Du}%
\affiliation{
CAS Key Laboratory of Microscale Magnetic Resonance and School of Physical Sciences, University of Science and Technology of China, Hefei 230026, China}
\affiliation{
CAS Center for Excellence in Quantum Information and Quantum Physics, University of Science and Technology of China, Hefei 230026, China}
\maketitle

\section{Construction of arbitrary quantum channels}

The super-operator $\mathcal{P}$ can be constructed by arbitrary
physical processes, which is described by CPTP (Completely Positive Trace
Preserving) map $\mathcal{M}$ and satisfies
\begin{equation}
    \left(\mathcal{M}\hat{\rho}\right)^{\dagger}=\mathcal{M}\hat{\rho},\quad 
        \mathcal{M}\hat{\rho}\geq 0,\quad 
        \Tr\mathcal{M}\hat{\rho}=\Tr\hat{\rho}.
\end{equation}
Here $ \mathcal{M}\hat{\rho} $ can always be expanded as
\begin{equation}
\mathcal{M}\hat{\rho}=\sum_{n,m=1}^{D}C_{n,m}\hat{A}_{n}\hat{\rho}\hat{A}_{m}^{\dagger},\label{eq:CPTP}
\end{equation}
where $D$ is the dimension of Hilbert space and $\hat{A}_{n}$ is
complete ($\sum_{n}\hat{A}_{n}\hat{A}_{n}^{\dagger}=\mathds{1}$) and orthogonal ($\Tr\hat{A}_{m}^{\dagger}\hat{A}_{n}=\delta_{mn}$) operators to expand the density matrix. For the case of spin-1/2 system ($D=2$),
there are four $\hat{A}_{n}$ which can be four Pauli matrices $\{\mathds{1},\hat{\sigma}_{x},\hat{\sigma}_{y},\hat{\sigma}_{z}\}$.
If $\mathcal{M}$ is CPTP, $C_{nm}$ must be Hermite ($C^{\dagger}=C$) and positive matrix ($C\ge 0$). The elements of the super-operator $\hat{A}_{n}\rho\hat{A}_{m}^{\dagger}$ can in principle be constructed by combination of a complete set of various $\mathcal{M}$ with different $C_{n,m}$. This indicates that the super-operator $\mathcal{P}$ can be constructed arbitrarily
through combination of a complete set of CPTP map $\mathcal{M}$.

\begin{table}
    \begin{tabular}{|c|c|c|}
    \hline 
    Symbol$ \backslash $ Operation & The physical meaning & Matrix form\tabularnewline
    \hline 
    \hline 
    $\mathcal{R}^{0}$ & rotation around any axis by angle $0$ & $\mathbf{R}(0,\mathbf{n})$\tabularnewline
    \hline 
    $\mathcal{R}_{x}^{\pi/2}$ & rotation around axis $x$ by angle $\pi/2$ & $\mathbf{R}(\pi/2,\mathbf{e}_{x})$\tabularnewline
    \hline 
    $\mathcal{R}_{y}^{\pi/2}$ & rotation around axis $y$ by angle $\pi/2$ & $\mathbf{R}(\pi/2,\mathbf{e}_{y})$\tabularnewline
    \hline 
    $\mathcal{R}_{z}^{\pi/2}$ & rotation around axis $z$ by angle $\pi/2$ & $\mathbf{R}(\pi/2,\mathbf{e}_{z})$\tabularnewline
    \hline 
    $\mathcal{R}_{x}^{-\pi/2}$ & rotation around axis $x$ by angle $-\pi/2$ & $\mathbf{R}(-\pi/2,\mathbf{e}_{x})$\tabularnewline
    \hline 
    $\mathcal{R}_{y}^{-\pi/2}$ & rotation around axis $y$ by angle $-\pi/2$ & $\mathbf{R}(-\pi/2,\mathbf{e}_{y})$\tabularnewline
    \hline 
    $\mathcal{R}_{z}^{-\pi/2}$ & rotation around axis $z$ by angle $-\pi/2$ & $\mathbf{R}(-\pi/2,\mathbf{e}_{z})$\tabularnewline
    \hline 
    $\mathcal{R}_{xy}^{\pi/2}$ & rotation around axis $\frac{x+y}{\sqrt{2}}$ by angle $\pi/2$ & $\mathbf{R}(\pi/2,\frac{\mathbf{e}_{x}+\mathbf{e}_{y}}{\sqrt{2}})$\tabularnewline
    \hline 
    $\mathcal{R}_{yz}^{\pi/2}$ & rotation around axis $\frac{y+z}{\sqrt{2}}$ by angle $\pi/2$ & $\mathbf{R}(\pi/2,\frac{\mathbf{e}_{y}+\mathbf{e}_{z}}{\sqrt{2}})$\tabularnewline
    \hline 
    $\mathcal{R}_{zx}^{\pi/2}$ & rotation around axis $\frac{z+x}{\sqrt{2}}$ by angle $\pi/2$ & $\mathbf{R}(\pi/2,\frac{\mathbf{e}_{z}+\mathbf{e}_{x}}{\sqrt{2}})$\tabularnewline
    \hline 
    \textcolor{red}{$\mathcal{M}_{x}$} & Measure along the $x$ axis & $\vert+x\rangle\langle+x\vert\hat{\rho}\vert+x\rangle\langle+x\vert-\vert-x\rangle\langle-x\vert\hat{\rho}\vert-x\rangle\langle-x\vert$\tabularnewline
    \hline 
    \textcolor{red}{$\mathcal{M}_{y}$} & Measure along the $y$ axis & $\vert+y\rangle\langle+y\vert\hat{\rho}\vert+y\rangle\langle+y\vert-\vert-y\rangle\langle-y\vert\hat{\rho}\vert-y\rangle\langle-y\vert$\tabularnewline
    \hline 
    \textcolor{red}{$\mathcal{M}_{z}$} & Measure along the $z$ axis & $\vert+z\rangle\langle+z\vert\hat{\rho}\vert+z\rangle\langle+z\vert-\vert-z\rangle\langle-z\vert\hat{\rho}\vert-z\rangle\langle-z\vert$\tabularnewline
    \hline 
    \textcolor{red}{$\mathcal{P}_{z}$} & Polarizing the state to $\vert-z\rangle$ & $\vert0\rangle\langle0\vert\rho\vert0\rangle\langle0\vert+\vert0\rangle\langle1\vert\rho\vert1\rangle\langle0\vert=\frac{1}{4}\left[\mathcal{J}_{0}-2\mathcal{J}_{z}^{+}+\left(\mathcal{J}_{xx}^{+}+\mathcal{J}_{yy}^{+}+\mathcal{J}_{zz}^{+}\right)-2\mathcal{J}_{xy}^{-}\right]\hat{\rho}$\tabularnewline
    \hline 
    \textcolor{red}{$\mathcal{P}_{x}$} & Polarizing the state to $\vert-x\rangle$ & $\mathcal{R}_{y}^{\pi/2}\mathcal{P}_{z}\mathcal{R}_{y}^{-\pi/2}\hat{\rho}=\frac{1}{4}\left[\mathcal{J}_{0}-2\mathcal{J}_{x}^{+}+\left(\mathcal{J}_{xx}^{+}+\mathcal{J}_{yy}^{+}+\Sigma_{zz}^{+}\right)-2\mathcal{J}_{yz}^{-}\right]\hat{\rho}$\tabularnewline
    \hline 
    \textcolor{red}{$\mathcal{P}_{y}$} & Polarizing the state to $\vert-y\rangle$ & $\mathcal{R}_{x}^{-\pi/2}\mathcal{P}_{z}\mathcal{R}_{x}^{\pi/2}\hat{\rho}=\frac{1}{4}\left[\mathcal{J}_{0}-2\mathcal{J}_{y}^{+}+\left(\mathcal{J}_{xx}^{+}+\mathcal{J}_{yy}^{+}+\mathcal{J}_{zz}^{+}\right)-2\mathcal{J}_{zx}^{-}\right]\hat{\rho}$\tabularnewline
    \hline 
    \end{tabular}\caption{The definition of the sixteen operations. The red symbols indicate the operations beyond the NMR control. \label{tab:complete operations}}
\end{table}

Taking spin 1/2 system for example, there are sixteen basic super-operators
$\hat{\sigma}_{i}\hat{\rho}\hat{\sigma}_{j}$($i=0,x,y,z$). For convenience,
we regroup it into symmetric and anti-symmetric form
\begin{equation}
    \begin{aligned}
        \mathcal{J}_{0}&=\hat{\sigma}_{0}\hat{\rho}\hat{\sigma}_{0},\\
        \mathcal{J}_{\alpha}^{+}\hat{\rho}&=(\hat{\sigma}_{\alpha}\hat{\rho}+\hat{\rho}\hat{\sigma}_{\alpha})/2,\qquad\qquad\qquad\alpha=x,y,z,\\
        \mathcal{J}_{\alpha}^{-}\hat{\rho}&=-\ii(\hat{\sigma}_{\alpha}\hat{\rho}-\hat{\rho}\hat{\sigma}_{\alpha})/2,\qquad\qquad\qquad\alpha=x,y,z,\\
        \mathcal{J}_{\alpha\beta}^{+}&=\mathcal{J}_{\beta\alpha}^{+}=(\hat{\sigma}_{\alpha}\hat{\rho}\hat{\sigma}_{\beta}+\hat{\sigma}_{\beta}\hat{\rho}\hat{\sigma}_{\alpha})/2,\qquad\alpha<\beta=x,y,z,\\
        \mathcal{J}_{\alpha\beta}^{-}&=-\mathcal{J}_{\beta\alpha}^{-}=-\ii(\hat{\sigma}_{\alpha}\hat{\rho}\hat{\sigma}_{\beta}-\hat{\sigma}_{\beta}\hat{\rho}\hat{\sigma}_{\alpha})/2,\qquad\alpha<\beta=x,y,z.
    \end{aligned}\label{eq:elements}
\end{equation}
We consider how to construct these generators through the physical
process. Firstly, we consider the unitary evolution. Unitary evolution
can be written as $\hat{U}=\cos\theta\hat{\sigma}_{0}-\ii\sin\theta n_{\alpha}\hat{\sigma}_{\alpha}$
\begin{equation}
    \begin{aligned}
        \hat{U}\hat{\rho}\hat{U}^{\dagger}&=\left(\cos\theta\hat{\sigma}_{0}-\ii\sin\theta n_{\alpha}\hat{\sigma}_{\alpha}\right)\hat{\rho}\left(\cos\theta\hat{\sigma}_{0}+\ii\sin\theta n_{\alpha}\hat{\sigma}_{\alpha}\right)\\
        &=\cos^{2}\theta\hat{\sigma}_{0}\hat{\rho}\hat{\sigma}_{0}+2\cos\theta\sin\theta n_{\alpha}\frac{-\ii\left(\hat{\sigma}_{\alpha}\hat{\rho}\hat{\sigma}_{0}-\hat{\sigma}_{0}\hat{\rho}\hat{\sigma}_{\alpha}\right)}{2}+\sin^{2}\theta n_{\alpha}n_{\beta}\frac{\hat{\sigma}_{\alpha}\hat{\rho}\hat{\sigma}_{\beta}+\hat{\sigma}_{\beta}\hat{\rho}\hat{\sigma}_{\alpha}}{2}\\
        &=\left[\cos^{2}\theta\mathcal{J}_{0}+2\cos\theta\sin\theta n_{\alpha}\mathcal{J}_{\alpha}^{-}+\sin^{2}\theta n_{\alpha}n_{\beta}\mathcal{J}_{\alpha\beta}^{+}\right]\hat{\rho}.
    \end{aligned}
\end{equation}
The multiply algebra of $\mathcal{J}_{0}$, $\mathcal{J}_{\alpha}^{-}$
and $\mathcal{J}_{\alpha\beta}^{+}$ is self-closing and is not enough
to generate the elements $\mathcal{J}_{\alpha}^{+},\mathcal{J}_{\alpha\beta}^{-}$. This means that unitary evolution is not complete to generate all
the generator in equation~(\eqref{eq:elements}).

If dissipation process is introduced, the generator can be constructed
completely. For example, if we can do measurement of $\hat{\sigma}_{z}$
, the density matrix will collapse to $\hat{\rho}\rightarrow\vert\pm z\rangle\langle\pm z\vert$
with probability $p_{\pm}=\langle\pm z\vert\hat{\rho}\vert\pm z\rangle$.
Then we can construct the process
\begin{equation}
\hat{\rho}\rightarrow\sum_{u=\pm}p_{u}\vert uz\rangle\langle uz\vert\equiv\mathcal{M}_{z}\hat{\rho}=\vert+z\rangle\langle+z\vert\hat{\rho}\vert+z\rangle\langle+z\vert-\vert-z\rangle\langle-z\vert\hat{\rho}\vert-z\rangle\langle-z\vert.\label{eq:Mz}
\end{equation}
Using $\mathcal{M}_{z}$, we can construct the process $\mathcal{J}_{z}^{+}$.
Combined with the unitary rotation, $\mathcal{J}_{x\slash y}^{+}$
can also be generated. The element can be introduced by polarization
process as shown in \eqref{tab:complete operations}. As a result,
all the sixteen basic elements can be generated via the set of complete
operations shown in \eqref{tab:complete operations} (Note: these
operations is not unique). 

Since $\mathbf{\mathbf{G}}^{\prime}=\{\mathcal{R}^{0},\mathcal{R}_{x/y/z}^{\pm\pi/2},\mathcal{R}_{xy}^{\pi},\mathcal{R}_{yz}^{\pi},\mathcal{R}_{zx}^{\pi},\mathcal{M}_{x/y/z}\mathcal{P}_{x/y/z}\}$ constitute a complete but non-orthogonal basis for any quantum channel $\mathcal{P}$, then our aim is the construction of $\mathcal{P}$ though the above set of physical operations $\mathbf{\mathbf{G}}^{\prime}$, namely, 
\begin{equation}
    \mathcal{P}=\sum_{i=1}^{16}p_{i}\mathrm{G}_{i}^{\prime}.
\end{equation}
The central work is the calculation of the weight $p_{i}$. We discuss
the problem in the basis of $1,\hat{\sigma}_{x},\hat{\sigma}_{y},\hat{\sigma}_{z}$ and the matrix form of super-operator is denoted by blackboard bold
\begin{equation}
    \mathbb{P}=\sum_{i}p_{i}\mathrm{\mathbb{G}}_{i}^{\prime}.
\end{equation}
Then we project the above equation in the basis $\{\mathbb{G}_{n+4(m-1)}=\ketbra{m}{n},1\le n,m\le4\}$,
which is complete $\sum_{i}\mathbb{G}_{i}\mathbb{G}_{i}^{\dagger}=1$
and orthogonal $\Tr\mathbb{G}_{i}^{\dagger}\mathbb{G}_{j}=\delta_{ij}$. For example 
\begin{equation}
    \mathbb{G}_{2}=
    \begin{pmatrix}0 & 1 & 0 & 0\\
        0 & 0 & 0 & 0\\
        0 & 0 & 0 & 0\\
        0 & 0 & 0 & 0
    \end{pmatrix},
\end{equation}
so we find
\begin{equation}
    \mathbf{A}=\mathbf{T}\cdot\mathbf{p},
\end{equation}
where $T_{ji}=\Tr\mathrm{\mathbb{G}}_{j}^{\dagger}\mathrm{\mathbb{G}}_{i}^{\prime}$\textcolor{blue}{{}
}and $A_{j}=\Tr\mathrm{\mathbb{G}}_{j}^{\dagger}\cdot\mathbb{P}$.
The weight is calculated to be
\begin{equation}
    \mathbf{p}=\mathbf{T}^{-1}\cdot\mathbf{A}.
\end{equation}
For example, the quantum channel $\mathbb{P}_{xx}=\ketbra{2}{2}$
can be expanded to
\begin{equation}
    \begin{aligned}
        \mathbb{P}_{xx}&=\begin{pmatrix}0 & 0 & 0 & 0\\
            0 & 1 & 0 & 0\\
            0 & 0 & 0 & 0\\
            0 & 0 & 0 & 0
            \end{pmatrix}=\frac{1}{2}\mathbb{R}^{0}+\frac{1}{4}\mathbb{R}_{x}^{\pi/2}-\frac{1}{4}\mathbb{R}_{y}^{\pi/2}-\frac{1}{4}\mathbb{R}_{z}^{\pi/2}+\frac{1}{4}\mathbb{R}_{x}^{-\pi/2}-\frac{1}{4}\mathbb{R}_{y}^{-\pi/2}-\frac{1}{4}\mathbb{R}_{z}^{-\pi/2}\\
            &=\frac{1}{2}\mathbb{R}^{0}+\frac{1}{4}\left(\mathbb{R}_{x}^{\pi/2}+\mathbb{R}_{x}^{-\pi/2}\right)-\frac{1}{4}\left(\mathbb{R}_{y}^{\pi/2}+\mathbb{R}_{y}^{-\pi/2}\right)-\frac{1}{4}\left(\mathbb{R}_{z}^{\pi/2}+\mathbb{R}_{z}^{-\pi/2}\right).
    \end{aligned}
\end{equation}
The weight $p_{i}$ for other sparse quantum channel has been calculated
in \eqref{tab:construction}.

\begin{table}
    \begin{tabular}{|c|c|c|c|c|c|c|c|c|c|c|c|c|c|c|c|c|}
    \hline 
    sparse elements$ \backslash $ operation & $\mathcal{R}^{0}$ & $\mathcal{R}_{x}^{\pi/2}$ & $\mathcal{R}_{y}^{\pi/2}$ & $\mathcal{R}_{z}^{\pi/2}$ & $\mathcal{R}_{x}^{-\pi/2}$ & $\mathcal{R}_{y}^{-\pi/2}$ & $\mathcal{R}_{z}^{-\pi/2}$ & $\mathcal{R}_{xy}^{\pi}$ & $\mathcal{R}_{yz}^{\pi}$ & $\mathcal{R}_{zx}^{\pi}$ & \textcolor{blue}{$\mathcal{M}_{x}$} & \textcolor{blue}{$\mathcal{M}_{y}$} & \textcolor{blue}{$\mathcal{M}_{z}$} & \textcolor{blue}{$\mathcal{P}_{x}$} & \textcolor{blue}{$\mathcal{P}_{y}$} & \textcolor{blue}{$\mathcal{P}_{z}$}\tabularnewline
    \hline 
    $\mathcal{P}_{00}$ & $-\frac{1}{2}$ & $\frac{1}{4}$ & $\frac{1}{4}$ & $\frac{1}{4}$ & $\frac{1}{4}$ & $\frac{1}{4}$ & $\frac{1}{4}$ & $0$ & $0$ & $0$ & \textcolor{blue}{$0$} & \textcolor{blue}{$0$} & \textcolor{blue}{$0$} & \textcolor{blue}{$0$} & \textcolor{blue}{$0$} & \textcolor{blue}{$0$}\tabularnewline
    \hline 
    \textcolor{magenta}{$\mathcal{P}_{0x}$} & $\frac{1}{2}$ & $-\frac{1}{4}$ & $-\frac{1}{4}$ & $-\frac{1}{4}$ & $-\frac{1}{4}$ & $-\frac{1}{4}$ & $-\frac{1}{4}$ & $0$ & $0$ & $0$ & \textcolor{blue}{$1$} & \textcolor{blue}{$0$} & \textcolor{blue}{$0$} & \textcolor{blue}{$1$} & \textcolor{blue}{$0$} & \textcolor{blue}{$0$}\tabularnewline
    \hline 
    \textcolor{magenta}{$\mathcal{P}_{0y}$} & $\frac{1}{2}$ & $-\frac{1}{4}$ & $-\frac{1}{4}$ & $-\frac{1}{4}$ & $-\frac{1}{4}$ & $-\frac{1}{4}$ & $-\frac{1}{4}$ & $0$ & $0$ & $0$ & \textcolor{blue}{$0$} & \textcolor{blue}{$1$} & \textcolor{blue}{$0$} & \textcolor{blue}{$0$} & \textcolor{blue}{$1$} & \textcolor{blue}{$0$}\tabularnewline
    \hline 
    \textcolor{magenta}{$\mathcal{P}_{0z}$} & $\frac{1}{2}$ & $-\frac{1}{4}$ & $-\frac{1}{4}$ & $-\frac{1}{4}$ & $-\frac{1}{4}$ & $-\frac{1}{4}$ & $-\frac{1}{4}$ & $0$ & $0$ & $0$ & \textcolor{blue}{$0$} & \textcolor{blue}{$0$} & \textcolor{blue}{$1$} & \textcolor{blue}{$0$} & \textcolor{blue}{$0$} & \textcolor{blue}{$1$}\tabularnewline
    \hline 
    \textcolor{magenta}{$\mathcal{P}_{x0}$} & $-\frac{1}{2}$ & $\frac{1}{4}$ & $\frac{1}{4}$ & $\frac{1}{4}$ & $\frac{1}{4}$ & $\frac{1}{4}$ & $\frac{1}{4}$ & $0$ & $0$ & $0$ & \textcolor{blue}{$0$} & \textcolor{blue}{$0$} & \textcolor{blue}{$0$} & \textcolor{blue}{$-1$} & \textcolor{blue}{$0$} & \textcolor{blue}{$0$}\tabularnewline
    \hline 
    $\mathcal{P}_{xx}$ & $\frac{1}{2}$ & $\frac{1}{4}$ & $-\frac{1}{4}$ & $-\frac{1}{4}$ & $\frac{1}{4}$ & $-\frac{1}{4}$ & $-\frac{1}{4}$ & $0$ & $0$ & $0$ & \textcolor{blue}{$0$} & \textcolor{blue}{$0$} & \textcolor{blue}{$0$} & \textcolor{blue}{$0$} & \textcolor{blue}{$0$} & \textcolor{blue}{$0$}\tabularnewline
    \hline 
    $\mathcal{P}_{xy}$ & $\frac{1}{2}$ & $-\frac{1}{4}$ & $-\frac{1}{4}$ & $-\frac{1}{4}$ & $-\frac{1}{4}$ & $-\frac{1}{4}$ & $\frac{1}{4}$ & $\frac{1}{2}$ & $0$ & $0$ & \textcolor{blue}{$0$} & \textcolor{blue}{$0$} & \textcolor{blue}{$0$} & \textcolor{blue}{$0$} & \textcolor{blue}{$0$} & \textcolor{blue}{$0$}\tabularnewline
    \hline 
    $\mathcal{P}_{xz}$ & $\frac{1}{2}$ & $-\frac{1}{4}$ & $\frac{1}{4}$ & $-\frac{1}{4}$ & $-\frac{1}{4}$ & $-\frac{1}{4}$ & $-\frac{1}{4}$ & $0$ & $0$ & $\frac{1}{2}$ & \textcolor{blue}{$0$} & \textcolor{blue}{$0$} & \textcolor{blue}{$0$} & \textcolor{blue}{$0$} & \textcolor{blue}{$0$} & \textcolor{blue}{$0$}\tabularnewline
    \hline 
    \textcolor{magenta}{$\mathcal{P}_{y0}$} & $-\frac{1}{2}$ & $\frac{1}{4}$ & $\frac{1}{4}$ & $\frac{1}{4}$ & $\frac{1}{4}$ & $\frac{1}{4}$ & $\frac{1}{4}$ & $0$ & $0$ & $0$ & \textcolor{blue}{$0$} & \textcolor{blue}{$0$} & \textcolor{blue}{$0$} & \textcolor{blue}{$0$} & \textcolor{blue}{$-1$} & \textcolor{blue}{$0$}\tabularnewline
    \hline 
    $\mathcal{P}_{yx}$ & $\frac{1}{2}$ & $-\frac{1}{4}$ & $-\frac{1}{4}$ & $\frac{1}{4}$ & $-\frac{1}{4}$ & $-\frac{1}{4}$ & $-\frac{1}{4}$ & $\frac{1}{2}$ & $0$ & $0$ & \textcolor{blue}{$0$} & \textcolor{blue}{$0$} & \textcolor{blue}{$0$} & \textcolor{blue}{$0$} & \textcolor{blue}{$0$} & \textcolor{blue}{$0$}\tabularnewline
    \hline 
    $\mathcal{P}_{yy}$ & $\frac{1}{2}$ & $-\frac{1}{4}$ & $\frac{1}{4}$ & $-\frac{1}{4}$ & $-\frac{1}{4}$ & $\frac{1}{4}$ & $-\frac{1}{4}$ & $0$ & $0$ & $0$ & \textcolor{blue}{$0$} & \textcolor{blue}{$0$} & \textcolor{blue}{$0$} & \textcolor{blue}{$0$} & \textcolor{blue}{$0$} & \textcolor{blue}{$0$}\tabularnewline
    \hline 
    $\mathcal{P}_{yz}$ & $\frac{1}{2}$ & $-\frac{1}{4}$ & $-\frac{1}{4}$ & $-\frac{1}{4}$ & $\frac{1}{4}$ & $-\frac{1}{4}$ & $-\frac{1}{4}$ & $0$ & $\frac{1}{2}$ & $0$ & \textcolor{blue}{$0$} & \textcolor{blue}{$0$} & \textcolor{blue}{$0$} & \textcolor{blue}{$0$} & \textcolor{blue}{$0$} & \textcolor{blue}{$0$}\tabularnewline
    \hline 
    \textcolor{magenta}{$\mathcal{P}_{z0}$} & $-\frac{1}{2}$ & $\frac{1}{4}$ & $\frac{1}{4}$ & $\frac{1}{4}$ & $\frac{1}{4}$ & $\frac{1}{4}$ & $\frac{1}{4}$ & $0$ & $0$ & $0$ & \textcolor{blue}{$0$} & \textcolor{blue}{$0$} & \textcolor{blue}{$0$} & \textcolor{blue}{$0$} & \textcolor{blue}{$0$} & \textcolor{blue}{$-1$}\tabularnewline
    \hline 
    $\mathcal{P}_{zx}$ & $\frac{1}{2}$ & $-\frac{1}{4}$ & $-\frac{1}{4}$ & $-\frac{1}{4}$ & $-\frac{1}{4}$ & $\frac{1}{4}$ & $-\frac{1}{4}$ & $0$ & $0$ & $\frac{1}{2}$ & \textcolor{blue}{$0$} & \textcolor{blue}{$0$} & \textcolor{blue}{$0$} & \textcolor{blue}{$0$} & \textcolor{blue}{$0$} & \textcolor{blue}{$0$}\tabularnewline
    \hline 
    $\mathcal{P}_{zy}$ & $\frac{1}{2}$ & $\frac{1}{4}$ & $-\frac{1}{4}$ & $-\frac{1}{4}$ & $-\frac{1}{4}$ & $-\frac{1}{4}$ & $-\frac{1}{4}$ & $0$ & $\frac{1}{2}$ & $0$ & \textcolor{blue}{$0$} & \textcolor{blue}{$0$} & \textcolor{blue}{$0$} & \textcolor{blue}{$0$} & \textcolor{blue}{$0$} & \textcolor{blue}{$0$}\tabularnewline
    \hline 
    $\mathcal{P}_{zz}$ & $\frac{1}{2}$ & $-\frac{1}{4}$ & $-\frac{1}{4}$ & $\frac{1}{4}$ & $-\frac{1}{4}$ & $-\frac{1}{4}$ & $\frac{1}{4}$ & $0$ & $0$ & $0$ & \textcolor{blue}{$0$} & \textcolor{blue}{$0$} & \textcolor{blue}{$0$} & \textcolor{blue}{$0$} & \textcolor{blue}{$0$} & \textcolor{blue}{$0$}\tabularnewline
    \hline 
    \end{tabular}\caption{Expansion of the basic sparse matrix on the representation of sixteen
    operations. The red operations can not be accessible by unitary operations
    because it need the incoherent operations as indicated by blue color.
    \label{tab:construction}}
\end{table}

\section{Extraction of arbitrary correlations in the case
of general interaction}

The protocol in the main text can also been generalized to extracting the correlations for general interaction form as following
\begin{equation}
    \hat{V}(t)=\sum_{\alpha=1}^{d}\hat{S}_{\alpha}(t)\otimes\hat{B}_{\alpha}(t).
\end{equation}
We consider the weighted signal 
\begin{equation}
    S_{N}=\sum_{\boldsymbol{\beta}_{N}=1}^{D^{4}}p_{\boldsymbol{\beta}_{N}}S_{N,\boldsymbol{\beta}_{N}},\label{eq:signal_general}
\end{equation}
where $\boldsymbol{\beta}_{N}=\{\beta_{N},\cdot\cdot\cdot,\beta_{2},\beta_{1}\}$
is the index collective. $ p_{\boldsymbol{\beta}_{N}}$ is the weight for signal $S_{N,\boldsymbol{\beta}_{N}}$ defined as
\begin{equation}
    S_{N,\boldsymbol{\beta}_{N}}=\mathrm{Tr}\left[\hat{O}\mathcal{U}_{N}\mathcal{C}_{\alpha_{N}}\cdot\cdot\cdot\mathcal{C}_{\alpha_{i+1}}\mathcal{U}_{i}\mathcal{C}_{\alpha_{i}}\cdot\cdot\cdot\mathcal{C}_{\alpha_{2}}\mathcal{U}_{1}\mathcal{C}_{\alpha_{1}}\left(\hat{\text{\ensuremath{\rho}}}_{S}\otimes\hat{\rho}_{\mathrm{B}}\right)\right],
\end{equation}
where each of the quantum channel is assigned to one element of the complete basis $ \mathcal{C}_{\alpha}(\alpha=1,\cdot\cdot\cdot,D^{4}) $
for quantum channel ($D$ is the dimension of Hilbert space of sensor). 

We expand $\mathcal{U}_{i}$ to the first order of
$\delta t$ and find that
\begin{equation}
    \mathcal{U}_{i}=\mathcal{T}e^{\int_{t_{i}}^{t_{i}+\delta t}\mathcal{L}(t)dt}=\sum_{\alpha_{i},\eta_{i}}\left(2\delta t\right)^{\vert\eta_{i}\vert}\mathcal{S}_{\alpha_{i}}^{\overline{\eta_{i}}}\mathcal{B}_{\alpha_{i}}^{\eta_{i}},
\end{equation}
where $\eta_{i}=0,\pm1$ and $\mathcal{S}_{\alpha_{i}}^{\overline{\eta}_{i}}$ are defined as
\begin{equation}
    \mathcal{S}_{\alpha_{i}}^{\overline{\eta}_{i}}=
    \begin{cases}
        1 & \eta_{i}=0,\\
        \mathcal{S}_{\alpha_{i}}^{-/+} & \eta_{i}=+\slash-.
    \end{cases}
\end{equation}
If $\eta_{k}=0$, the index $\alpha_{k}$ takes no value and it denotes that the sensor is assigned to the super-operator $1$. If $\eta_{k}=+\slash-$,
the index $\alpha_{i}$ takes the values $1,2,\cdot\cdot\cdot,d$. Then the signal $S_{N}$ becomes
\begin{equation}
    S_{N}=\sum_{\boldsymbol{\beta}_{N}=1}^{D^{4}}p_{\boldsymbol{\beta}_{N}}\sum_{\boldsymbol{\eta}_{N},\boldsymbol{\alpha}_{N}}\left(2\delta t\right)^{\sum_{i=1}^{N}\vert\eta_{i}\vert}A_{\boldsymbol{\alpha}_{N},\boldsymbol{\beta}_{N}}^{\overline{\boldsymbol{\eta}}_{N}}C_{\boldsymbol{\alpha}_{N}}^{\boldsymbol{\eta}_{N}}+O(\delta t^{N+1}).
\end{equation}
$C_{\boldsymbol{\alpha}_{N}}^{\boldsymbol{\eta}_{N}}\equiv C_{\alpha_{N}\cdot\cdot\cdot\alpha_{1}}^{\eta_{N}\cdot\cdot\cdot\eta_{1}}$ is the correlation defined as
\begin{equation}
    C_{\boldsymbol{\alpha}_{N}}^{\boldsymbol{\eta}_{N}}=\mathrm{Tr}_{\mathrm{B}}\left[\mathcal{B}_{\alpha_{N}}^{\eta_{N}}(t_{N})\cdot\cdot\cdot\mathcal{B}_{\alpha_{2}}^{\overline{\eta}_{2}}(t_{2})\mathcal{B}_{\alpha_{1}}^{\overline{\eta}_{1}}(t_{1})\hat{\text{\ensuremath{\rho}}}_{\mathrm{B}}\right],
\end{equation}
where
\begin{equation*}
    \mathcal{B}_{\alpha_{k}}^{\eta_{k}}(t_{k})=
    \begin{cases}
        1 & \eta_{k}=0,\\
        \mathcal{B}_{\alpha_{k}}^{+/-}(t_{k}) & \eta_{k}=+\slash-.
    \end{cases}
\end{equation*}
The index follows the same convention as $\mathcal{S}_{\alpha_{k}}^{\overline{\eta}_{k}}(t_{k})$. If there is one index in $\eta_{1},\eta_{2},\cdot\cdot\cdot,\eta_{N}$
being $0$, $C_{\alpha_{N}\cdot\cdot\cdot\alpha_{1}}^{\eta_{N}\cdot\cdot\cdot\eta_{1}}$ is a correlation with order less than $N$. For example, $C_{12\cdot3}^{+-0+}=\mathrm{Tr}_{\mathrm{B}}\left[\mathcal{B}_{1}^{+}(t_{4})\mathcal{B}_{2}^{-}(t_{3})1\mathcal{B}_{3}^{+}(t_{1})\hat{\text{\ensuremath{\rho}}}_{\mathrm{B}}\right]=\mathrm{Tr}_{\mathrm{B}}\left[\mathcal{B}_{1}^{+}(t_{4})\mathcal{B}_{2}^{-}(t_{3})\mathcal{B}_{3}^{+}(t_{1})\hat{\text{\ensuremath{\rho}}}_{\mathrm{B}}\right]$
is the third order correlation. 

After summation over $\boldsymbol{\beta}_{N}$, the signal is simplified to be
\begin{equation}
    S_{N}=\sum_{\boldsymbol{\eta}_{N},\boldsymbol{\alpha}_{N}}\left(2\delta t\right)^{\sum_{i=1}^{N}\vert\eta_{i}\vert}A_{\boldsymbol{\alpha}_{N}}^{\overline{\boldsymbol{\eta}}_{N}}C_{\boldsymbol{\alpha}_{N}}^{\boldsymbol{\eta}_{N}}+O(\delta t^{N+1}),
\end{equation}
where
\begin{equation}
    A_{\boldsymbol{\alpha}_{N}}^{\overline{\boldsymbol{\eta}}_{N}}=\sum_{\boldsymbol{\beta}_{N}=1}^{D^{4}}p_{\boldsymbol{\beta}_{N}}A_{\boldsymbol{\alpha}_{N},\boldsymbol{\beta}_{N}}^{\overline{\boldsymbol{\eta}}_{N}}\label{eq:coeA}
\end{equation}
and $A_{\boldsymbol{\alpha}_{N},\boldsymbol{\beta}_{N}}^{\overline{\boldsymbol{\eta}}_{N}}$ is the coefficient
\begin{equation}
    A_{\boldsymbol{\alpha}_{N},\boldsymbol{\beta}_{N}}^{\overline{\boldsymbol{\eta}}_{N}}=\mathrm{Tr}_{\mathrm{S}}\left[\hat{O}\mathcal{S}_{\alpha_{N}}^{\overline{\eta_{N}}}\mathcal{C}_{\beta_{N}}\cdot\cdot\cdot\mathcal{C}_{\beta_{i+1}}\mathcal{S}_{\alpha_{i}}^{\overline{\eta_{i}}}\mathcal{C}_{\beta_{i}}\cdot\cdot\cdot\mathcal{C}_{\beta_{2}}\mathcal{S}_{\alpha_{1}}^{\overline{\eta_{1}}}\mathcal{C}_{\beta_{1}}\hat{\text{\ensuremath{\rho}}}_{\mathrm{S}}\right].
\end{equation}
It is a constant if the observable $\hat{O}$ and $\hat{\rho}_{\mathrm{S}}$ are fixed.

If we can design the weight $p_{\boldsymbol{\beta}_{N}}$
to make the coefficient $A_{\boldsymbol{\alpha}_{N}}^{\overline{\boldsymbol{\eta}}_{N}}$ [see equation~\eqref{eq:coeA}] satisfies
\begin{equation}
    A_{\boldsymbol{\alpha}_{N}}^{\overline{\boldsymbol{\eta}}_{N}}\propto
    \begin{cases}
        1 & \boldsymbol{\eta}_{N}=\boldsymbol{\eta}^{0}_{N},\boldsymbol{\alpha}_{N}=\boldsymbol{\boldsymbol{\alpha}}^{0}_{N},\\
        0 & {\rm else}.
    \end{cases}\label{eq:cond}
\end{equation}
Since some correlations $C_{\boldsymbol{\alpha}_{N}}^{\boldsymbol{\eta}_{N}}$
vanish naturally because the trace of commutator always gives vanishing
result (for example, the correlations $C_{\alpha_{N}\alpha_{N-1}\cdot\cdot\cdot\alpha_{1}}^{-\eta_{N-1}\cdot\cdot\cdot\eta_{1}}, C_{\alpha_{N}\alpha_{N-1}\cdot\cdot\cdot\alpha_{1}}^{0-\cdot\cdot\cdot\eta_{1}}, \cdot\cdot\cdot, C_{\alpha_{N}\alpha_{N-1}\cdot\cdot\cdot\alpha_{2}\alpha_{1}}^{00\cdot\cdot\cdot0-}$), the coefficients $A_{\boldsymbol{\alpha}_{N}}^{\overline{\boldsymbol{\eta}}_{N}}$
associated with these correlations are irrelavant. The number of these
coefficients is
\begin{equation}
    d\left(2d\right)^{N-1}+d\left(2d\right)^{N-2}+\cdot\cdot\cdot d\left(2d\right)^{0}\equiv\frac{(2d)^{N+1}-d}{2d-1}.
\end{equation}
When these coefficients are excluded, the number of equations in equation~\eqref{eq:cond} is $(2d+1)^{N}-\frac{(2d)^{N+1}-d}{2d-1}$. There
are totally $D^{4N}$ for the control parameters $p_{\boldsymbol{\beta}_{N}}$
in equation~\eqref{eq:coeA}. If $D^{4}\ge2d+1$, then the solution of
equation~\eqref{eq:cond} always exists. For the spin Hamiltonian, $d=3$
and $D\ge2$ always satisfie the condition $D^{4}\ge 2d+1$ and hence the extraction of any correlation is always possible.

The signal equation~\eqref{eq:signal_general} will reduce
to the special case in the main text if $\eta_{\alpha_{N},\alpha_{N-1},\cdot\cdot\cdot,\alpha_{2},\alpha_{1}}$ takes the following form
\begin{equation}
    p_{\boldsymbol{\beta}_{N}}=\prod_{i=1}^{N}p_{\alpha_{i}}.
\end{equation}
By inserting it into equation~\eqref{eq:signal_general}, we find
\begin{equation}
    S_{N}=\mathrm{Tr}\left[\hat{O}\mathcal{U}_{\mathrm{I}}^{(N)}\mathcal{P}_{N}\cdot\cdot\cdot\mathcal{P}_{i+1}\mathcal{U}_{\mathrm{I}}^{(i)}\mathcal{P}_{i}\cdot\cdot\cdot\mathcal{P}_{2}\mathcal{U}_{\mathrm{I}}^{(1)}\mathcal{P}_{1}\left(\hat{\text{\ensuremath{\rho}}}_{S}\otimes\hat{\rho}_{\mathrm{B}}\right)\right],\label{eq:signal-1-2}
\end{equation}
where
\begin{equation}
    \mathcal{P}_{i}=\sum_{\alpha_{i}=1}^{D^{4}}p_{\alpha_{i}}\mathcal{C}_{\alpha_{i}}
\end{equation}
are the quantum channels used in the main text.

\section{Experimental errors and data analysis}
The error sources in the experiments include the control error caused by the $\pi/2$-pulse imperfection ($ E^{\pi/2} $), the evolution error of the quantum target caused by the RF inhomogeneity ($ E^{\rm evo} $), and the experimental readout error ($ E^{\rm r} $). After calibrating the deviation of $\pi/2$-pulse, the decay rate of the oscillatory signals, as well as the spectral fitting uncertainty, these noise impact can be investigated separately by numerically simulating the relative deviation between the signals $ \vec{x}_{\rm e} $ with only one error source and the signals $ \vec{x} $ without any error. The contributions of these error sources are presented in Supplementary Table 3 and 4.

\begin{figure*}
    \includegraphics[scale=0.9]{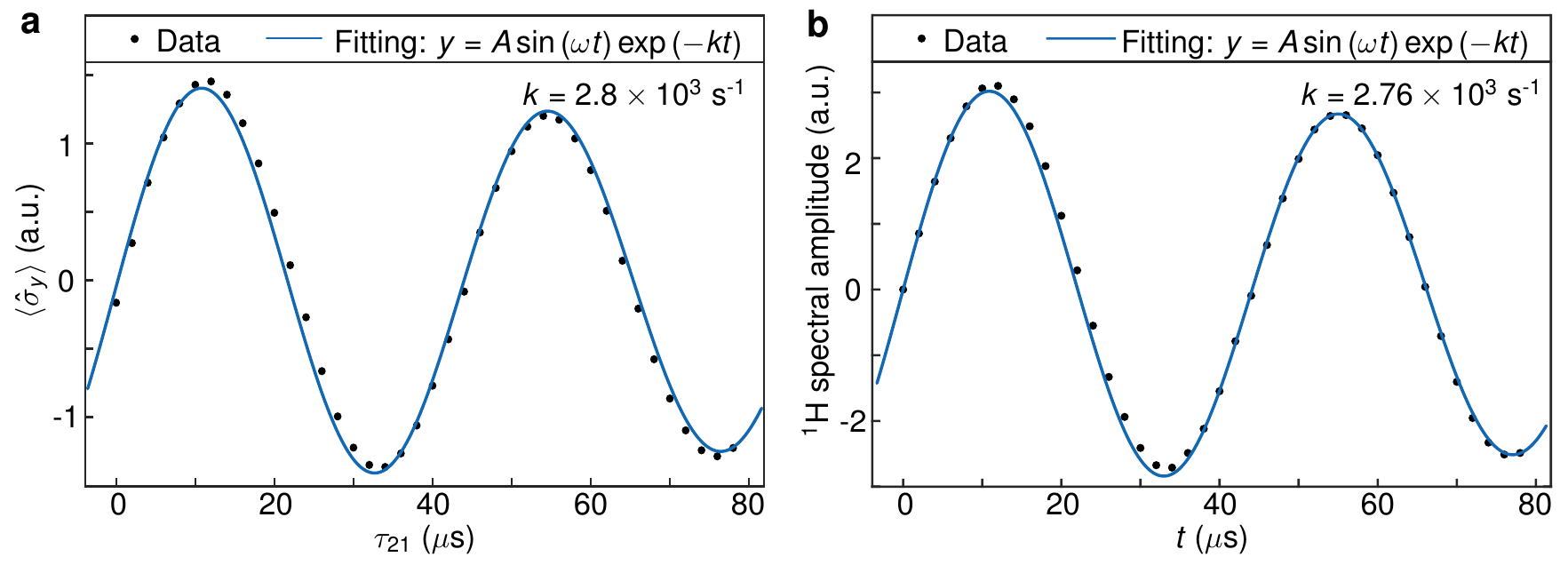}
    \caption{\textbf{The amplitude decay caused by the RF inhomogeneity. a}, The experimental data of the second-order correlation and the fitting curve. The fitting result is $ y\sim\sin\left(1.4\times 10^{5}t\right)\exp\left(-2.8\times 10^{3}t\right) $
    \textbf{b}, the NMR nutation spectroscopy of \textsuperscript{1}H and the fitting curve. The fitting result is $ y\sim\sin\left(1.4\times 10^{5}t\right)\exp\left(-2.76\times 10^{3}t\right) $.}\label{fig-decay}
\end{figure*}

It is worth noting that the second and fourth order correlations we measured in experiment both decay much faster than the $ T_{2} $ relaxation rate. Suppose that the signal amplitudes decay exponentially, i.e., $ A=A_{0}\ee^{-kt} $. By fitting the experimental data, we get the decay rate of the second-order correlation: $ k=2.8\times 10^{3} $ s\textsuperscript{-1} (see \ref{fig-decay}a), corresponding to the mean lifetime $ \tau=1/k=4\times 10^{-4} $ s. In contrast, the $ T_{2} $ relaxation time of the quantum target in our experiment, i.e., the protons in acetic acid, is typically about 0.3 s, which is much longer than $ \tau $. Thus the $ T_{2} $ relaxation must not be the main cause of the rapid decay. It is indeed caused by the inhomogeneity of the radio-frequency field ($ B_{1} $). To further demonstrate this, we directly measured the NMR nutation spectroscopy of \textsuperscript{1}H as shown in~\ref{fig-decay}b. By changing the length ($ t $) of the RF pulse along $ y $ axis, we obtain the integral values of the corresponding thermal equilibrium \textsuperscript{1}H spectra of different flip angles. The fitting result tells that the decay rate of the nutation spectroscopy is $ k=2.76\times 10^{3} $ s\textsuperscript{-1}, which is about the same rate as the second-order correlation.

\begin{table}
    \centering
    \begin{tabular}{p{1cm}p{2cm}p{2cm}p{2cm}p{2cm}p{2cm}p{2cm}}
    \hline 
    \hline 
    $ \delta t $ (ms) & $ \Delta^{\rm tot} $ & $ \Delta^{\rm th} $ & $ \Delta^{\rm exp} $ & $ \Delta^{\rm r} $ & $ \Delta^{\rm evo} $ & $ \Delta^{\pi/2} $ \\
    \hline 
    0.1 & 32.3\% & 0.2\%  & 32.3\% & 7.5\% & 13.1\% & 0.0\% \\
    0.5 & 18.8\% & 5.4\%  & 15.8\% & 0.4\% & 13.1\% & 0.1\%\\
      1 & 29.6\% & 20.2\%  & 15.3\% & 0.1\% & 13.1\% & 0.1\%\\
     \hline     
     \hline
    \end{tabular}
    \caption{Main deviations of second-order correlation}
\end{table}
\begin{table}
    \centering
    \begin{tabular}{p{2cm}p{2cm}p{2cm}p{2cm}p{2cm}p{2cm}p{2cm}p{2cm}}
    \hline 
    \hline 
     & $ \Delta^{\rm tot} $ & $ \Delta^{\rm th} $ & $ \Delta^{\rm exp} $ & $ \Delta^{\rm r} $ & $ \Delta^{\rm evo} $ & $ \Delta^{\pi/2} $ \\
    \hline 
    Robust & 23.5\% & 10.5\%  & 18.3\% & 1.0\% & 16.0\% & 0.4\% \\
    Non-robust & 112.2\% & 10.5\%  & 121.5\% & 2.6\% & 16.0\% & 109.1\% \\
     \hline     
     \hline
    \end{tabular}
    \caption{Main deviations of fourth-order correlation}\label{4th}
\end{table}

The radio-frequency field ($ B_{1} $) that varies along the NMR tube axis ($ z $ axis) is assumed to be the main cause of the RF inhomogeneity~\cite{Smith2001,Michaeli2008}. To theoretically describe the quantum dynamics under the inhomogeneous pulse, we suppose the strength of radio-frequency field applied on the sample is a function of the $ z $ coordinate: $ B\left(z\right) $ (in Hz), and the coordinate range of the sample column is $ -L\leq z\leq L $. Then the general state evolution under the inhomogeneous RF pulse is
\begin{equation}
    \hat{\rho}\left(t\right)=\int_{-L}^{L}\dd z\ee^{-\ii 2\pi B\left(z\right)\sigma_{y}t}\hat{\rho}_{0}\ee^{\ii 2\pi B\left(z\right)t}/\Tr\left[\int_{-L}^{L}\dd z\ee^{-\ii 2\pi B\left(z\right)t}\hat{\rho}_{0}\ee^{\ii 2\pi B\left(z\right)\sigma_{y}t}\right],\label{inho}
\end{equation}
where $ \hat{\rho}_{0} $ is the initial state of the \textsuperscript{1}H spin ensemble. The representation is actually the same as the state evolution of the phase damping process, which is a nonunitary map of the quantum state and is determined by the field distribution along the $ z $ axis. A typical model is that the rotation angle $ \omega=2\pi B\left(z\right) $ represented as a random variable which has a Gaussian distribution with mean $ \lambda $ and variance $ s $. Then the final state from this process is given by the density matrix obtained from averaging over $ \theta $:
\begin{equation}
    \hat{\rho}_{\rm f}=\frac{1}{\sqrt{2\pi s}}\int_{-\infty}^{\infty}\dd\omega\exp{-\frac{\left(\omega-\lambda\right)^{2}}{2 s}}\ee^{-\ii\sigma_{y}t\omega/2}\hat{\rho}_{\rm i}\ee^{\ii\sigma_{y}t\omega/2}.
\end{equation}
If $ \hat{\rho}_{\rm i}=\left(\mathds{1}+\epsilon\sigma_{z}\right)/2 $, using the Gaussian integral formulas:
\begin{equation}
    \frac{1}{\sqrt{2\pi s}}\int_{-\infty}^{\infty}\dd x\sin x\exp{-\frac{\left(x-x_{0}\right)^{2}}{2 s}}=\sin x_{0}\ee^{-s/2},\quad\frac{1}{\sqrt{2\pi s}}\int_{-\infty}^{\infty}\dd x\cos x\exp{-\frac{\left(x-x_{0}\right)^{2}}{2 s}}=\cos x_{0}\ee^{-s/2},
\end{equation}
we have
\begin{equation}
    \hat{\rho}_{\rm f}=\frac{1}{2}
    \begin{pmatrix}
        1+\epsilon\cos\left(\lambda t\right)\ee^{-st/2} & \epsilon\sin\left(\lambda t\right)\ee^{-st/2}\\
        \epsilon\sin\left(\lambda t\right)\ee^{-st/2} & 1-\epsilon\cos\left(\lambda t\right)\ee^{-st/2}
    \end{pmatrix}.
\end{equation}
Therefore the quantum coherence terms of the density matrix decay exponentially to zero with time. If using the Gaussian model in our experiment, we have $ s\approx 5\times 10^{3} $ s\textsuperscript{-1}. 

\begin{figure*}
    \includegraphics[scale=1]{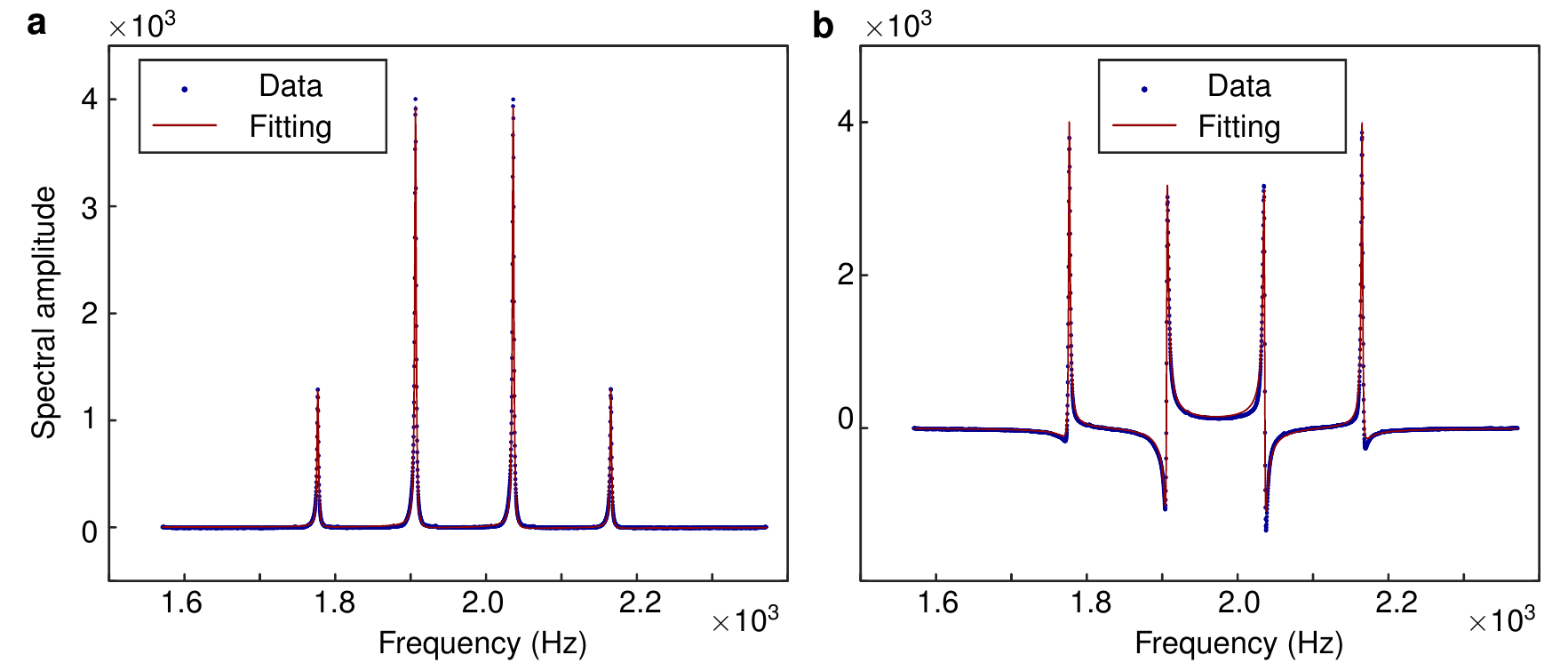}
    \caption{\textbf{The typical spectra from the experiments and their fitting curves. a}, The thermal equilibrium spectrum of $^{13}\mathrm{C}$. \textbf{b}, One of the spectra of the 2nd-order correlation measurement. The curve fitting equation is the Lorentzian function.}\label{fig-data}
\end{figure*}

\begin{figure*}
    \includegraphics[scale=1]{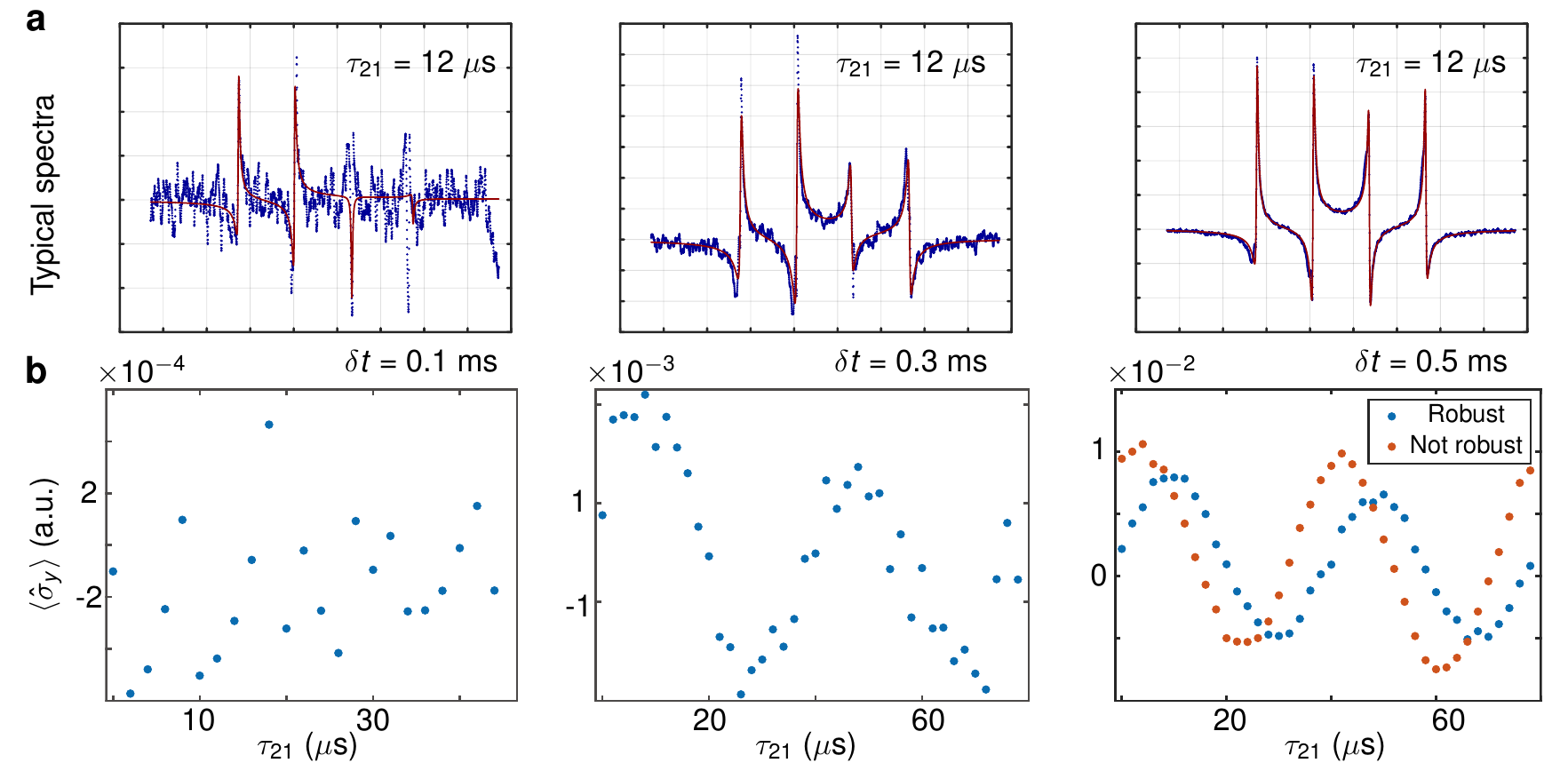}
    \caption{\textbf{Additional experimental data of the 4th-order correlation measurement. a}, The signal-to-noise ratios (SNR) of the typical spectra of different $ \delta t $. \textbf{b}, The corresponding 4th-order correlation signals in different $ \delta t $.}\label{fig-4th}
\end{figure*}

Besides the RF inhomogeneity, there are other errors in our experiments including the decoherence, the error of $ \pi/2 $ pulse and the readout error. The coherence time of the sample is $ T_{2}\approx 0.3 $ s, which is much longer than the time scale of the quantum correlations whose period is $ T\approx 40 $ $ \mu $s. So the errors from $ T_{2} $ decoherence are negligible compared to the errors contributed by the RF inhomogeneity. By fitting the NMR nutation spectroscopy, the relative error of $ \pi/2 $ pulse is characterized ($ 2\%\sim 3\% $), which directly reduces the fidelity of the quantum channels. As analyzed in the main text (including Methods), the pulse error has a great impact on the extraction of high order correlations because of the lower-order signal leakage, and we have used the error-resilient quantum channel by repeating the non-ideal for $ n $ times to suppress the unwanted low-order correlation. The readout error in the experiments refers to the fitting uncertainty of the NMR spectrum that determine the signals of quantum correlations. As shown in \ref{fig-data} and \ref{fig-4th}, by fitting the spectrum with Lorentzian function, the spectral integration corresponding to the expectation value $ \left<\sigma_{y}\right> $ is obtained. The fitting uncertainty, which is solely determined by the absolute intensity of the spectral ground noise (white noise), can be obtained by using the cftool toolbox in Matlab. The error bars presented in the figures of main experimental results are also from the fitting uncertainty.

\section{Optimal coupling-evolution time}
As discussed in the main text, there is a trade-off between the theoretical approximation and the SNR of the measured signals resulting from finite value of $ \delta t $. Here we discuss in detail about the trade-off and numerically find the optimal evolution time both for the measurement of second- and fourth-order correlations. 
The trade-off between the theoretical approximation and the SNR of the measured signals resulting from $ \delta t $ can be described by the total relative error of the target correlation signals:
\begin{equation}
    \Delta^{\rm tot}\left(\delta t\right)=\frac{\norm{\vec{E}^{\rm th}\left(\delta t\right)}+\norm{\vec{E}^{\rm exp}\left(\delta t\right)}}{\norm{\vec{S}^{\rm target}\left(\delta t\right)}}.\label{totE}
\end{equation}
Here $ \vec{E}^{\rm th}=\left|\vec{S}^{\rm sim}-\vec{S}^{\rm target}\right| $ is the theoretical approximation error from finite $ \delta t $. As defined in the main text, the target signals $ \vec{S}^{\rm target} $ for second- and fourth-order correlations are $ \delta t^2 C^{+-} $ and $ \delta t^4 p_{\rm C}C^{+--+} $ respectively. $ \vec{S}^{\rm sim} $ is the simulated signals of $ S_{2} $ or $ S_{4} $. $ \vec{E}^{\rm exp} $ is the total absolute deviation caused by the three main experimental error sources discussed above, i.e.,
\begin{equation}
    \norm{\vec{E}^{\rm exp}}=\norm{\vec{E}^{\pi/2}}+\norm{\vec{E}^{\rm evo}}+\norm{\vec{E}^{\rm r}},
\end{equation}
where $ \vec{E}^{\pi/2} $, $ \vec{E}^{\rm evo} $ and $ \vec{E}^{\rm r} $ present the control error caused by $\pi/2$-pulse imperfection, the evolution error of the quantum target caused by RF inhomogeneity, and the experimental readout error, respectively. With the error parameters measured in experiment (see Supplementary Note 3), all of these error mechanisms can be simulated individually. 

Here is the simulation process of each error mechanism:
\begin{itemize}
    \item $ \vec{E}^{\rm th} $: The theoretical approximation error can be simulated from the analytical formulas of second- and fourth-order correlations without any approximation:
    \begin{equation}
        \begin{aligned}
            S_{2}&=-\ii \Tr_{\rm B}\left\{\sin\left[\delta t \mathcal{B}^+\left(t_{2}\right)\right]\sin\left[\ii\delta t \mathcal{B}^-\left(t_{1}\right)\right]\hat{\rho}_{\rm B}\right\}\\
            S_{4}&=-p_{\rm C}\Tr\left\{\sin\left[\delta t \mathcal{B}^+\left(t_{4}\right)\sin\left[\ii\delta t \mathcal{B}^-\left(t_{3}\right)\right]\sin\left[\ii\delta t \mathcal{B}^-\left(t_{2}\right)\right]\sin\left[\delta t \mathcal{B}^+\left(t_{1}\right)\right]\right]\hat{\rho}_{\rm B}\right\}
        \end{aligned}
    \end{equation}
    Define $ \mathcal{B}^{\pm}_{k}\equiv\mathcal{B}^{\pm}\left(t_{k}\right) $, then by Taylor expansion we get the high order errors 
    \begin{equation}
        \begin{aligned}
            \vec{E}^{\rm th}_{\rm 2nd}&=\left|\vec{S}_{2}-\vec{S}^{\rm target}_{2}\right|\\
            &=\frac{1}{6}\delta t^4\left[\Tr_{\rm B}\left(\mathcal{B}^+_{2}\mathcal{B}^+_{2}\mathcal{B}^+_{2}\mathcal{B}^-_{1}\hat{\rho}_{\rm B}\right)-\Tr_{\rm B}\left(\mathcal{B}^+_{2}\mathcal{B}^-_{1}\mathcal{B}^-_{1}\mathcal{B}^-_{1}\hat{\rho}_{\rm B}\right)\right]+O\left(\delta t^6\right),\\
            \vec{E}^{\rm th}_{\rm 4th}&= \left|\vec{S}_{4}-\vec{S}^{\rm target}_{4}\right|\\
            &=\frac{p_{\rm C}}{6}\delta t^6\Big[\Tr_{\rm B}\left(\mathcal{B}^+_{4}\mathcal{B}^+_{4}\mathcal{B}^+_{4}\mathcal{B}^-_{3}\mathcal{B}^-_{2}\mathcal{B}^+_{1}\hat{\rho}_{\rm B}\right)-\Tr_{\rm B}\left(\mathcal{B}^+_{4}\mathcal{B}^-_{3}\mathcal{B}^-_{3}\mathcal{B}^-_{3}\mathcal{B}^-_{2}\mathcal{B}^+_{1}\hat{\rho}_{\rm B}\right)\\
            &\qquad\qquad\quad-\Tr_{\rm B}\left(\mathcal{B}^+_{4}\mathcal{B}^-_{3}\mathcal{B}^-_{2}\mathcal{B}^-_{2}\mathcal{B}^-_{2}\mathcal{B}^+_{1}\hat{\rho}_{\rm B}\right)+\Tr_{\rm B}\left(\mathcal{B}^+_{4}\mathcal{B}^-_{3}\mathcal{B}^-_{2}\mathcal{B}^+_{1}\mathcal{B}^+_{1}\mathcal{B}^+_{1}\hat{\rho}_{\rm B}\right)\Big]+O\left(\delta t^8\right),
        \end{aligned}
    \end{equation}
    with which $ \vec{E}^{\rm th}_{\rm 2nd} $ and $ E\vec{E}^{\rm th}_{\rm 4th} $ can be obtained numerically.
    \item $ \vec{E}^{\pi/2} $: The control error caused by $\pi/2$-pulse imperfection is simulated by replacing the error-free channel matrices into the non-ideal channel matrices. The analytical signal formulas of the non-ideal channels are available in Methods ($ S_{2}^{E} $ and $ S_{4}^{E} $), where the angle deviation $ \delta\theta $ is determined in experiment. For the second-order case and the fourth-order case with robust channels, the relative error from $\pi/2$-pulse imperfection ($ \Delta^{\pi/2} $) is almost negligible ($ \sim 0.1\% $).
    \item $ \vec{E}^{\rm evo} $: The evolution error caused by RF inhomogeneity is simulated by adding an exponential decay coefficient $ \ee^{-kt} $ before the simulated signal $ S^{\rm sim} $, where $ t $ is the evolution time and $ k=2.76\times 10^{3} $ s\textsuperscript{-1} is the decay rate of the nutation spectroscopy. Obviously, for the $ \Theta-$th order correlations, $ \vec{E}^{\rm evo} $ is the error mainly dependent of $ \delta t^{\Theta} $. Thus, the relative error $ \Delta^{\rm evo} $ is independent of $ \delta t $.
    \item $ \vec{E}^{\rm r} $: The signal readout error is simulated by adding a random Gaussian fluctuation on the final signals. The mean is $ \mu=0 $ and the standard deviation $ \sigma $ is the fitting error determined by Matlab cftool toolbox. $ \vec{E}^{\rm r} $ is totally independent of $ \delta t $.
\end{itemize}
Then equation~\eqref{totE} becomes
\begin{widetext}
    \begin{equation}
    \begin{aligned}
        \Delta^{\rm tot}\left(\delta t\right)&=\frac{\norm{\vec{E}^{\rm th}\left(\delta t\right)}+\norm{\vec{E}^{\rm \pi/2}\left(\delta t\right)}+ \norm{\vec{E}^{\rm evo}\left(\delta t\right)}+\norm{\vec{E}^{\rm r}}}{\norm{\vec{S}^{\rm target}\left(\delta t\right)}}\\
        &\approx\frac{\delta t^{\Theta+2}\norm{\vec{C}^{\eta_{N+2}\cdots\eta_{1}}}+\norm{\vec{E}^{\rm r}}}{\delta t^{\Theta}\norm{A^{\overline{\eta}_{N}\cdots\overline{\eta}_{1}}\vec{C}^{\eta_{N}\cdots\eta_{1}}}}+\Delta^{\pi/2}\left(\delta\theta\right)+\Delta^{\rm evo}
    \end{aligned} 
\end{equation}
\end{widetext}
Here $ \vec{S}^{\rm target}\left(\delta t\right)=\delta t^{\Theta}A^{\overline{\eta}_{N}\cdots\overline{\eta}_{1}}\vec{C}^{\eta_{N}\cdots\eta_{1}} $ is the target signals of the desired correlations. $ \Delta^{\pi/2}\left(\delta\theta\right) $ is the relative error caused by the $ \pi/2 $-pulse imperfection ($\pi/2 \to \pi/2+\delta\theta$), which is derived from equation~(14) and (23) in Methods:
\begin{equation}
    \Delta^{\pi/2}\left(\delta\theta\right) \approx
    \begin{cases}
    \delta\theta^2,\qquad\qquad\qquad\qquad~~~ {\rm for~~}C^{+-}, \\
    p_{\rm C}\left[1-\left(1-\delta\theta^{2}/2\right)^{3}\right],\quad {\rm for~~}C^{+--+}. 
    \end{cases}
\end{equation}
Note that the lower-order leakage in Methods-equation~(23) is greatly suppressed when Methods-equation~(26) is satisfied. $ \Delta^{\rm evo} $ is the relative error caused by RF inhomogeneity, i.e.,
\begin{equation}
    \Delta^{\rm evo}=\frac{\norm{\left(1-\ee^{-k\vec{\tau}}\right)\vec{C}^{\eta_{N}\cdots\eta_{1}}}}{\norm{\vec{C}^{\eta_{N}\cdots\eta_{1}}}}.
\end{equation}
Here $ k=2.76\times 10^{3} $ denotes the decay rate of the free evolution of the quantum target and $ \vec{\tau} $ is the sampling time list. $ \vec{E}^{\rm r} $ is determined by the ground noise of spectra and totally independent of $ \delta t $. Therefore, the $ \pi/2 $-pulse imperfection and rf inhomogeneity together contribute a constant relative error. Then, by taking $\partial \Delta^{\rm tot}_{\Theta}\left(\delta t\right) / \partial\Delta t=0$, the optimal evolution time is obtained:
\begin{equation}
    \delta t_{\rm opt}\approx\left(\frac{2\norm{\vec{C}^{\eta_{N+2}\cdots\eta_{1}}}}{\Theta\norm{\vec{E}^{\rm r}}}\right)^{1/(\Theta+2)}.
\end{equation}

\begin{figure*}
    \includegraphics[scale=1]{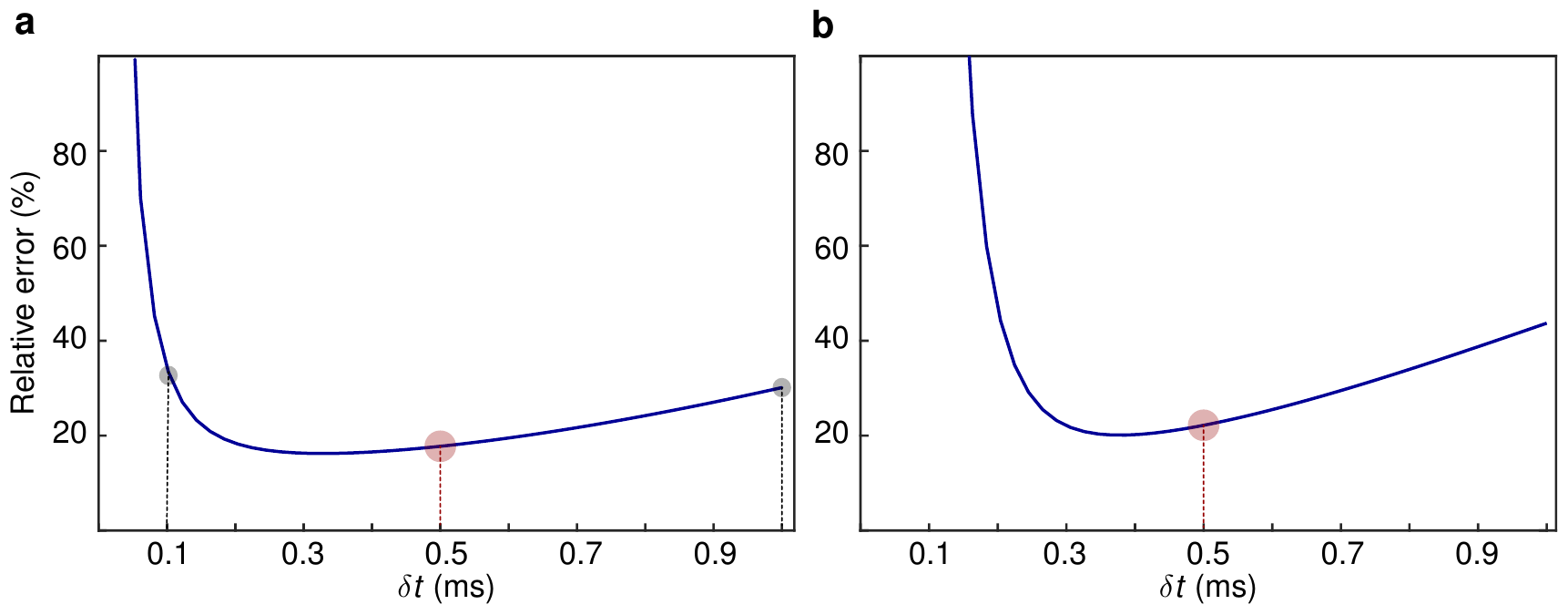}
    \caption{\textbf{Numerical simulation of the total relative error versus $ \delta t $. a}, The simulated relative measurement error of the second-order correlation versus $ \delta t $. \textbf{b}, The simulated relative measurement error of the fourth-order correlation versus $ \delta t $.}\label{fig-optimal}
\end{figure*}

The numerical simulation results of the total relative error versus $ \delta t $ are presented in \ref{fig-optimal}. \ref{fig-optimal}a displays the simulated result of the relative error of the second-order correlation. As shown in the figure, the total error becomes relatively large because of the first term for the small $ \delta t $. And for a large $ \delta t $, the second term contributes a lot, which also leads to a large total error. Therefore, we can find that the optimal $ \delta t $ is around $ 0.35 $ ms, which has the smallest error, and the value $ \delta t=0.5 $ ms we choose in the experiment also corresponds to a relatively small error. \ref{fig-optimal}b simulates the relative error of the fourth-order correlation. As we can see in the figure, the total error has a similar trend with that of the second-order correlation, and the optimal $ \delta t $ with the smallest error also lies between around $ 0.4 $ ms. Our choice $ \delta t=0.5 $ ms in the experiment is also appropriate.

\clearpage
\begin{figure*}
    \includegraphics[scale=1]{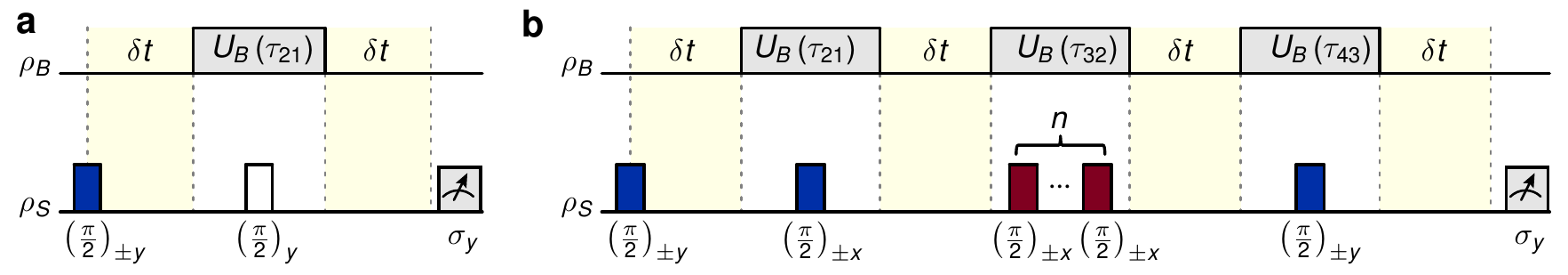}
    \caption{\textbf{The NMR pulse sequences for extracting the 2nd- and 4th-order correlations. a}, The pulse sequence of the second-order correlation measurement. 
    \textbf{b}, The pulse sequence of the fourth-order correlation measurement. The pulses labeled with $ \left(\pi/2\right)_{x/y} $ mean the $ \pi/2 $ rotation along the $ x/y $ axis, and the pulses labeled with $ \left(\pi/2\right)_{\pm x/y} $ mean the phase cycling $ \left(X+\bar{X}\right)/2 $ or $ \left(Y+\bar{Y}\right)/2 $. The shaded regions labeled with $ \delta t $ are the short time interaction processes.}\label{fig-24th}
\end{figure*}

\begin{figure*}
    \includegraphics[scale=1]{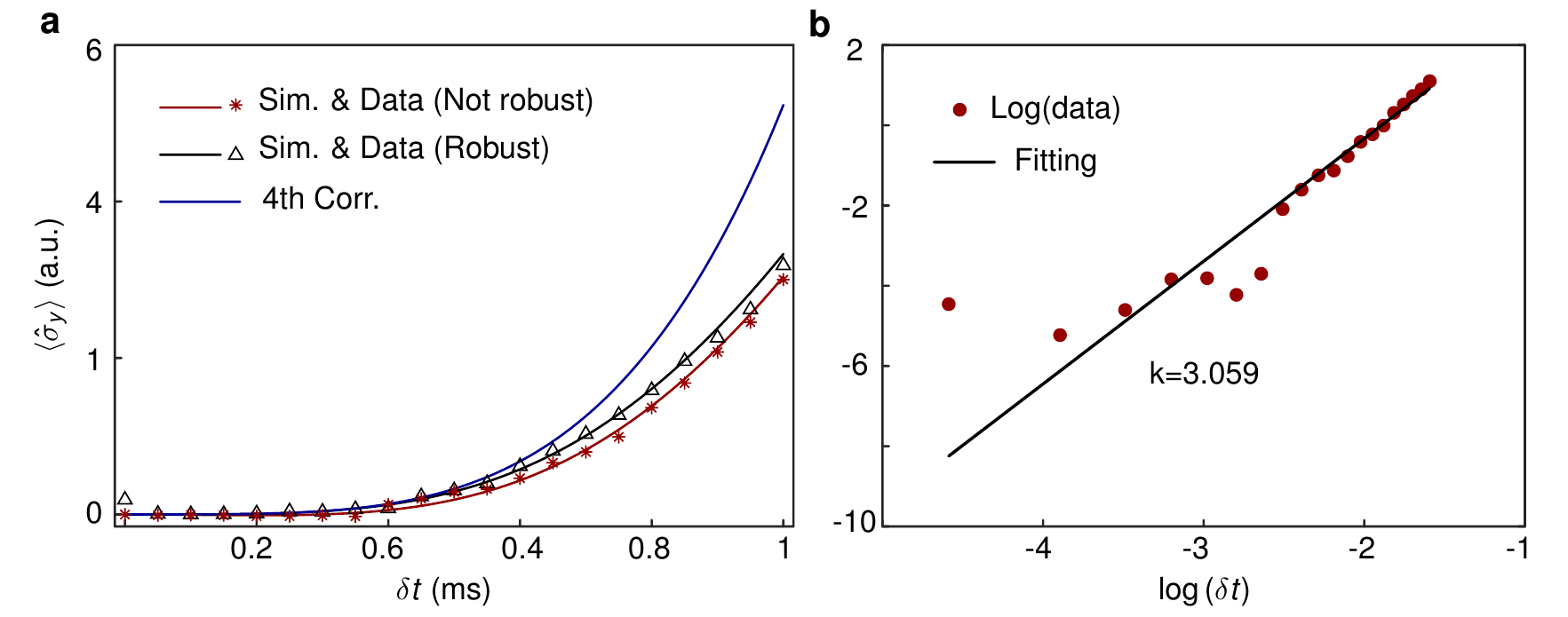}
    \caption{\textbf{Additional results of the non-robust 4th-order correlation measurement. a},
    The simulated result (red solid curve) and the experimental result (red stars) of the non-robust design versus $ \delta t $ compared with those of the robust design (black solid curve for the simulated result and black triangles for the experimental result) and the analytical result (solid blue curve). \textbf{b}, The corresponding fitting curve of the non-robust experimental result, which has a slope of $ 3.059 $. }\label{fig-4dt}
\end{figure*}



\clearpage